\pdfoutput=1
\documentclass[aps,prx,reprint,groupedaddress]{revtex4-2}

\usepackage[T1]{fontenc}
\usepackage[utf8]{inputenc}
\usepackage[english]{babel}

\usepackage{amsmath,amssymb}
\usepackage{autobreak} 

\newcommand*\diff{\mathop{}\!\mathrm{d}}

\usepackage{siunitx}
\DeclareSIUnit\Molar{\textsc{m}}

\usepackage{hyperref}
\hypersetup{
    colorlinks=true,
    linkcolor=blue,
    citecolor=blue,
    urlcolor=blue,
}

\usepackage{graphicx}
\usepackage{caption}

\graphicspath{{figures/}}

\begin{document}
\title{3D shape of epithelial cells on curved substrates}

\author{Nicolas Harmand}

\author{Sylvie Hénon}
\email{sylvie.henon@u-paris.fr}
\affiliation{Université de Paris, CNRS, Matière et Systèmes Complexes (MSC), UMR 7057, F-75006 Paris, France}

\date{\today}

\begin{abstract}

Epithelia are ubiquitous tissues found in plants and animals, in which cells closely bound to one another form continuous layers covering the surfaces of organs. 
They display a large diversity of functions and forms, from totally flat to highly curved.
Building a physical framework to account for the shape of cells in epithelia is thus an important challenge to understand various normal and pathological biological processes, such as epithelial embryogenesis or cancer metastasis.
It is widely recognized that the shape of epithelial cells is determined by the tension generated by the actomyosin cortex and the adhesion of cells to the substrate and to each other. 
These tensions and adhesions are not homogeneously distributed on the cell surface, which makes a 3D view of the problem valuable. 
To account for these biological and structural contributions to cell shape, different physical models have been proposed, which include surface energies, adhesions, line tensions, volume compressibility or elasticity terms.
However, an experimental procedure that would allow a validation of a minimal physical model for the shape of epithelial cells in 3D has not yet been proposed.
In this study, we cultured MDCK epithelial cells on substrates with a sinusoidal profile, allowing us to measure the shape of the cells on various positive and negative curvatures. 
We found that MDCK cells are thicker in the valleys than on the crests of sinusoidal substrates. 
The influence of curvature on the shape of epithelial cells could not be understood with a model using only differential apical, basal and lateral surface energies. 
However, the addition of an apical line tension was sufficient to quantitatively account for the experimental measurements. 
The model also accounts for the shape of MDCK cells that overexpress E-cadherin. 
On the other hand, when reducing myosin II activity with blebbistatin, we measured a saturation of the difference in cell thickness between valleys and crests, suggesting the need for a term limiting large cell deformations.
Our results show that a minimal model that accounts for epithelial cell shape needs to include an apical line tension in addition to differential surface energies, highlighting the importance of structures that produce anisotropic tension in epithelial cells, such as the actin belt linking adherens junctions. 
In the future, the model could be used to account for the shape of epithelial cells in different contexts, such as embryogenesis. 
Furthermore, our experimental procedure could be used to test a wider range of physical models for the shape of epithelia in curved environments, including, for example, continuous models.

\end{abstract}

\maketitle

\section{Introduction}

Physical understanding of the factors that regulate the shape of epithelial cells is important for various biological processes such as cancer \cite{wirtz_physics_2011} \cite{hanahan_hallmarks_2011} and embryogenesis \cite{dasgupta_physics_2018} \cite{mckinley_cellular_2018}. 
For instance, metastatic events, such as the so-called epithelial-mesenchymal transition or the invasion of neighboring tissues through the collective migration of epithelial cells, involve multiple cell shape changes.
Morphogenetic events that occur during embryogenesis also involve cell shape changes, in addition to division, apoptosis, growth and migration, in order to create higher order forms, which will become tissues and organs. 
These events often involve the occurrence of an out-of-plane curvature or a curvature inversion, that correlates with changes in the shape of the cells \cite{balaji_regulation_2018} \cite{Monier2015} \cite{fouchard_curling_2020}. 
The influence of curvature on the motility of epithelial cells has been studied recently \cite{xi_emergent_2017} \cite{yevick_architecture_2015} but little is known about its influence on the shape of epithelial cells. 
Epithelial tissues and liquid foams share a common morphological characteristic: they are cellular materials. 
It has thus been proposed that the shape of epithelial cells could be governed by surface energies, similarly to that of soap bubbles \cite{Thompson1945}. 
Differential cell-cell and cell-medium surface energies were also used to explain cell sorting \cite{steinberg_reconstruction_1963}. 
Various models from the family of vertex models were then proposed to understand the shape of epithelial cells in 2D, either in the plane of the epithelium \cite{farhadifar_influence_2007} \cite{landsberg_increased_2009} \cite{aegerter-wilmsen_integrating_2012} \cite{Bi2015}, or in cross-section \cite{Wen2017} \cite{polyakov_passive_2014} \cite{Storgel2016}. 
Studying the shape of cells in 3D is especially useful in curved environments and, to this end, 3D vertex models have been proposed more recently \cite{Hannezo2014} \cite{honda_three-dimensional_2004} \cite{Misra2016}. 
In these models, the energy of the cell is expressed with different terms accounting for the multiple force generating processes and structures in the epithelial cell.
Most often, vertex models include surface tensions, with different values for the different interfaces of the epithelial cells: basal, apical and lateral.
Using different values for the different surface tensions is a way to account for the adhesion energy at the lateral surfaces and at the basal surface and the inhomogeneity in the contractility of the actomyosin cortex.
The models also usually include an apical perimeter line tension or line elasticity that accounts for the contribution of the actin belt that links the adherens junctions around the cell, and is located close to the apical surface.
It also often includes a volume compressibility, which requires two parameters on its own: the compressibility modulus and the target volume.
Finally, most models include additional terms, which vary widely between one model and another. 
All of this makes the number of parameters in the models large and thus their experimental validation difficult.
To study the shape of epithelial cells, \textit{in-vitro} models are widely used in order to reduce the variability observed \textit{in-vivo}.
The MDCK cells have the biological components that contribute to the shape  \cite{Dukes2011}.

Here we combine modeling with the measurement of the thickness of epithelial cells as a function of the substrate curvature in a controlled experimental device to validate a minimal model that accounts for the shape of epithelial cells. 
We build sinusoidally shaped substrates, allowing for positive and negative out of plane curvatures to be next to each other and measure the thickness of MDCK cells on such substrates. We show that for large curvatures – radius of curvature less than typically \SI{30}{\micro\meter} – the cells are thicker in valleys than on crests. 
We then develop a 3D energy model for the cells, inspired by existing vertex models \cite{Hannezo2014}, which includes apical, basal and lateral surface tensions, as well as an apical line tension, and derive the cell energy in the particular geometry of a curved substrate. 
We show that such a model accounts for our measurements with only one adjustable parameter, the ratio of apical line tension to lateral surface tension, which we thus measure to be equal to \SI{6.75}{\micro\meter}. 
On the contrary a model with surface tensions only cannot account for our measurements. 
We further validate the model with measurements on cells over-expressing the intercellular adhesion protein E-cadherin. 
In the case of cells with decreased contractility through exposure to blebbistatin, the thickness difference between valleys and crests saturates at high curvature, suggesting that an additional term limiting large cellular or nuclear deformation should be added to the model. 

\section{Shape of epithelial cells on curved substrates}
\label{section_shape}
\begin{figure}[h]
	\begin{center}
    \includegraphics[width=\columnwidth]{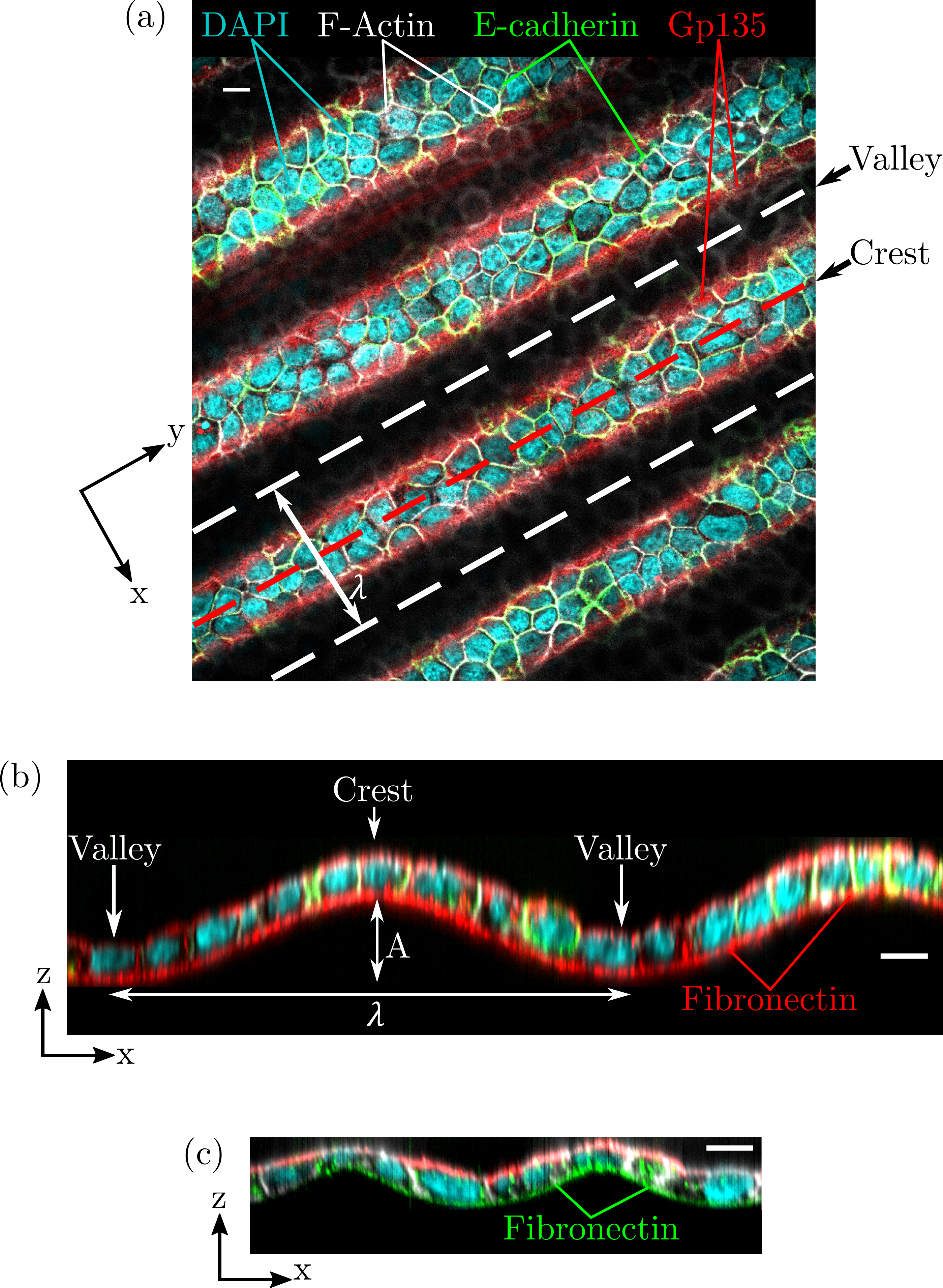}
	\caption{MDCK cells cultured on a substrate with a sinusoidal profile. The nucleus of the cells is cyan, F-actin is grey and the apical membrane is red.(a) Top view of EcadGFP MDCK cells, E-cadherin is green. (b) Cross-section view of EcadGFP MDCK cells grown on a sinusoidal substrate. E-cadherin is green and both, the apical membrane and fibronectin are red. (c) Cross-section view of WT MDCK cells grown on a sinusoidal substrate. Fibronectin is green. Scale bars = \SI{10}{\micro\meter}.}
	\label{fig_images_exp}
	\end{center}
\end{figure}

In order to study the influence of substrate curvature on their shape, epithelial cells were cultured on substrates with a sinusoidal profile.
The details of sample preparation, image acquisition and image analysis are described in appendix\ref{Materials and methods}.
Briefly, MDCK cells, either WT or stably expressing E-cadherin-GFP, were grown for 96h in total - and 48h after confluence - on PDMS substrates with a sinusoidal profile, coated with partly fluorescent fibronectin; cells were fixed and stained for DNA (DAPI), F-actin (SiR-actin) and apical membrane (anti-GP135); samples were finally imaged with a confocal microscope. 
Typical images are displayed in Fig. \ref{fig_images_exp}.

On substrates with a long wavelength $\lambda$ (Fig. \ref{fig_images_exp}(b)) the shape of the epithelium closely follows that of the substrate. On substrates with a short wavelength $\lambda$ on the contrary, the cells are thicker in the valleys than on the crests (Fig. \ref{fig_images_exp}(c)).
In order to report quantitatively this qualitative observation, we measured on cross-section images the thickness of the cells specifically located on the crests or in the valleys of the substrate (see Appendix \ref{analyse_images} for the details of the measurement). 
For each sample $i$ (i.e. here a microscope field of view, \SI{230}{\micro\meter} x \SI{230}{\micro\meter}), we obtained on average 59 cells in the valleys and 50 cells on the crests, and we measured the mean values of their heights $H^i_{valley}$ and $H^i_{crest}$, respectively. 
The relative height difference measured on sample $i$ is then:

\begin{equation}
\Delta h^i = 2\frac{H^i_{valley}-H^i_{crest}}{H^i_{valley}+H^i_{crest}}
\end{equation}

\begin{figure}[h]
	\centering
	\includegraphics[width=0.8\columnwidth]{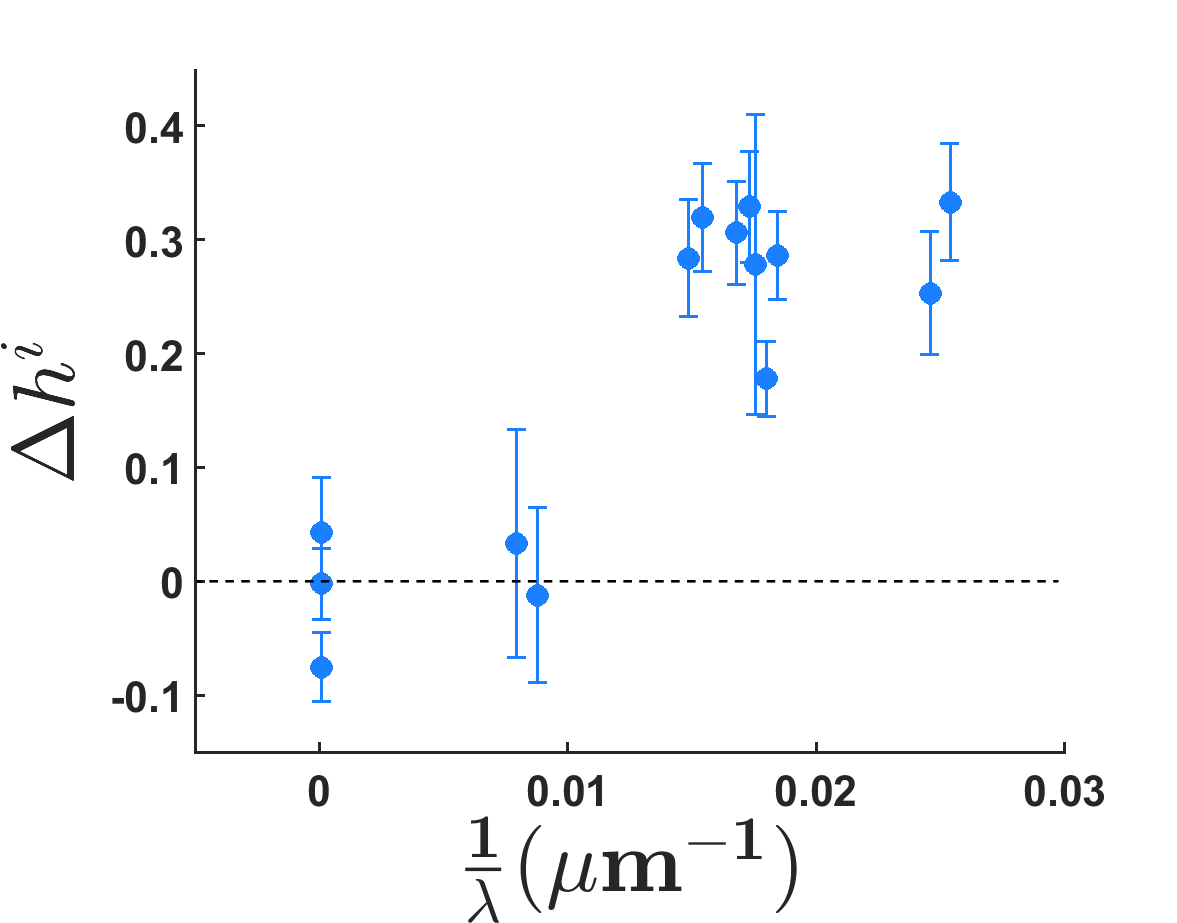}
	\caption{Relative height difference $\Delta h^i$ between cells on crests and cells in valleys $\Delta h^i$ as a function of the wave number of the substrate $\frac{1}{\lambda}$. Each data point corresponds to a microscope field of view on which the height of an average of 110 cells was measured. The error bars are the 95\% confidence intervals. Points at $\frac{1}{\lambda} = 0$ are control measurements on flat substrates.}
	\label{sinus_WT_k} 
\end{figure}

$\Delta h^i$ is therefore the difference in mean height between cells on crests and in valleys, relative to the mean height in sample $i$.
This choice of variable $\Delta h^i$ is made to smooth out the inter-sample variability of the mean cell thickness.
The measurements, performed on a total of 1873 cells from 14 fields of view of 6 independent experiments are plotted in Fig. \ref{sinus_WT_k}, as a function of the wave number $\frac{1}{\lambda}$. 

The error bars are the 95\% confidence intervals for each measurement. 
Data points at exactly zero wave number are controls for which the height of the cells on a flat substrate was measured by randomly positioning imaginary crests and valleys. This control shows that our measurement method does not create an artificial height difference between crests and valleys larger than $\pm 5\%$. 
This dispersion of $\pm 5\%$ is mainly explained by the spatial fluctuations in the cells thickness of a given sample.

The measured relative height differences $\Delta h^i$ are not significantly different from 0 for the largest wavelengths tested, up to $\frac{1}{\lambda} \simeq \SI{0.009}{\per\micro\meter}$. 
At lower wavelengths on the contrary, from $\frac{1}{\lambda} = \SI{0.014}{\per\micro\meter}$ to $\frac{1}{\lambda} = \SI{0.26}{\per\micro\meter}$, the measured height differences are about 30\%.

\section{Surface and line tensions model}
\label{section_model}
Our goal here is to propose a minimal model that can account for the shape of the cells on flat and sinusoidal substrates and, more specifically, to predict the height of the cells in the substrate valleys compared to the height of the cells on the crests.
These predictions will then be compared with the experimental results.

\subsection{Geometry of the problem}
\label{section_model_geometrie}
\begin{figure}[h]
	\begin{center}
		\includegraphics[width=0.6\linewidth]{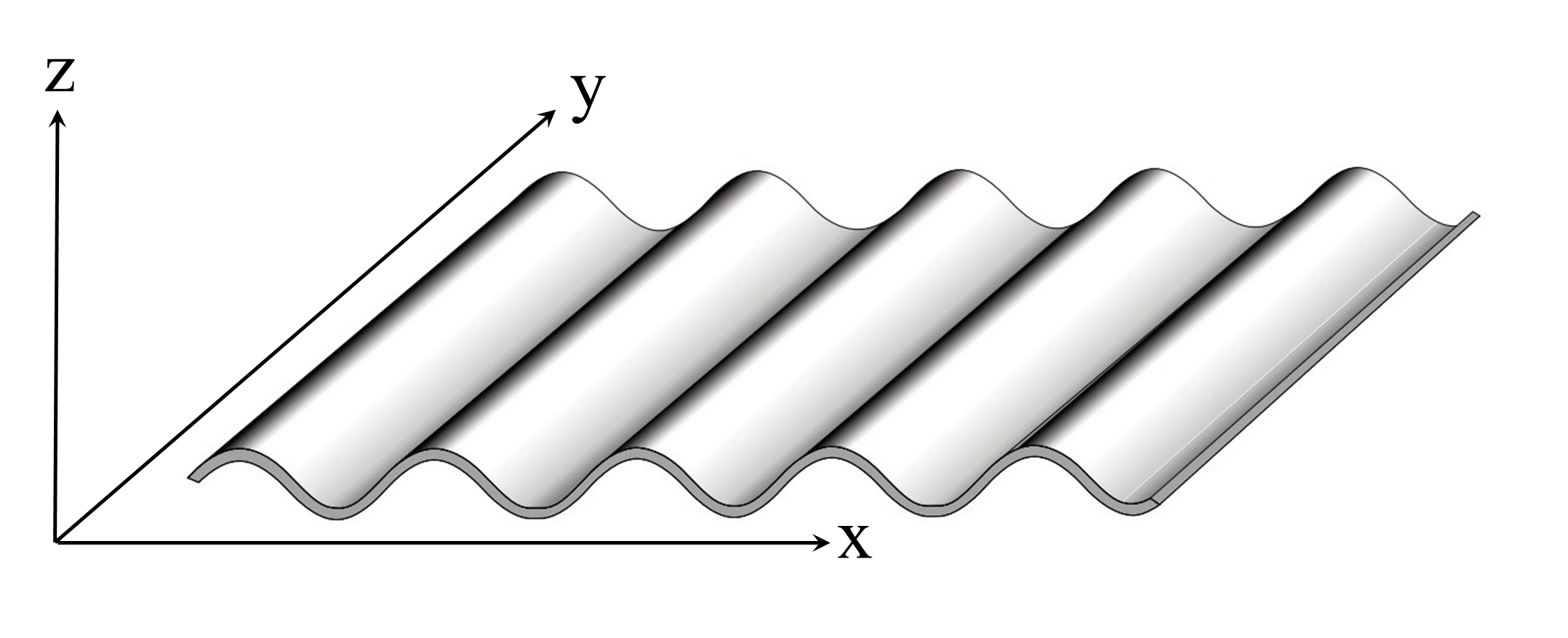}
	\end{center}
	\caption{Diagram of the substrate.}
	\label{tole_ondulee}
\end{figure}

The geometry in which the cells are placed is constrained by the shape of the substrates, schematized in Fig. \ref{tole_ondulee}.
In the $y$-direction, the substrate is invariant while it has a sinusoidal profile in the $x$-direction.
The $z$-direction is orthogonal to the mean plane of the substrate as shown in Fig. \ref{tole_ondulee} and Fig. \ref{schema_modele}.

In order to compute the shape of the cells on such a substrate, we make a few geometrical assumptions.
It is first assumed that the intercellular junctions are flat and orthogonal to the substrate.
It is also assumed that the apical surfaces are not curved.
Actually, they are curved towards the outside of the cells but these curvatures are small enough to consider the quantities of surface area and volume lost by this approximation negligible.

\subsubsection{Flat substrate}
\label{geometry_flat}

When the cells are on a flat substrate, their shape is invariant along the $z$-axis, since the intercellular junctions are assumed to be orthogonal to the substrate.
The apical surface area $S_a$ and basal surface area $S_b$ are therefore equal to each other.
They are also equal to $S_{\textsc{\tiny 1/2}}$, defined as the surface area of the cell, at its mid-point along the $z$-axis, in the $xy$-plane: $S_{a} = S_{b} = S_{\textsc{\tiny 1/2}}$, as depicted in Fig. \ref{schema_modele}(b).

The intercellular surface area $S_{cc}$ is related to $S_{\textsc{\tiny 1/2}}$ and to $H$, the thickness of the cell, \textit{i.e.} its length in the direction orthogonal to the substrate, through the shape index $\alpha$ of the mid-height surface $\alpha$: $S_{cc} = \alpha\sqrt{S_{\textsc{\tiny 1/2}}}H$, with $\alpha=\frac{P_{\textsc{\tiny 1/2}}}{\sqrt{S_{\textsc{\tiny 1/2}}}}$, $P_{\textsc{\tiny 1/2}}$ being the perimeter of the cell at mid-height.

\subsubsection{Curved substrate}
\label{geometry_curved}

The profile of the substrate is given by the relationship 
\begin{equation}
z(x) = \frac{A}{2}\sin \left( \frac{2\pi}{\lambda}x\right)
\end{equation}

The height and surface area of the cells now depend on their position along the sinusoid. 
They are noted $H(x)$ and $S_{\textsc{\tiny 1/2}}(x)$, respectively.

\begin{figure}[h]
	\begin{center}
    	\includegraphics[width=\columnwidth]{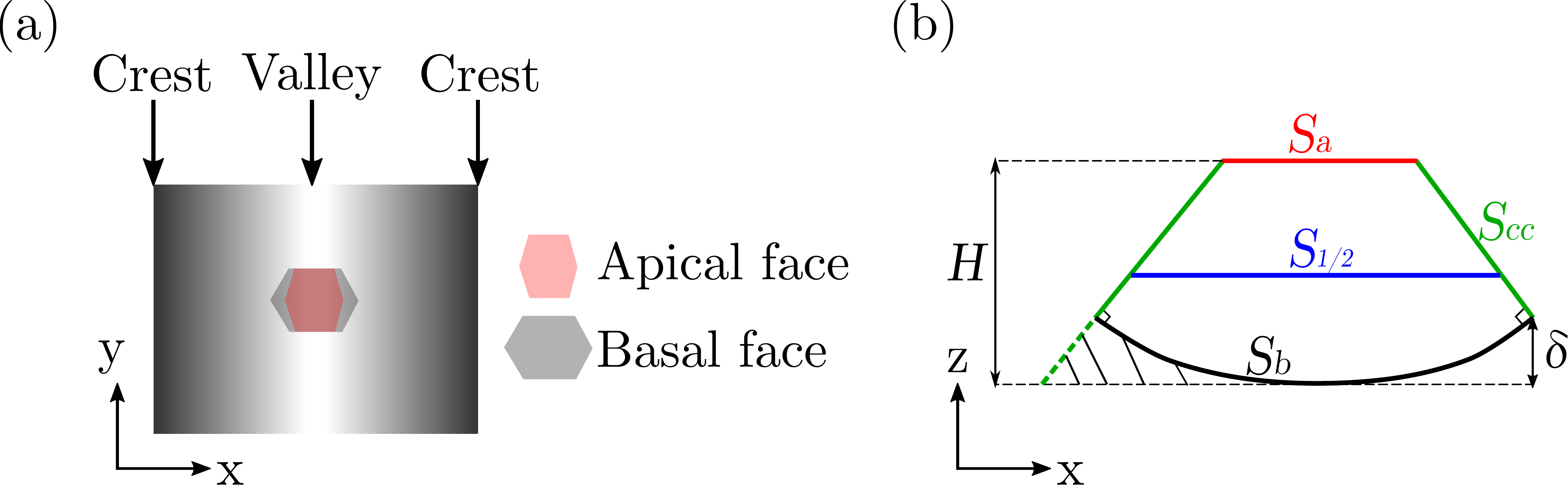}
    \end{center}
    \caption{Diagram of the model for a cell in a valley. (a) Top view of a cell in a valley. (b) Cross-section view in the direction orthogonal to the $y$-axis.}
    \label{schema_modele}
\end{figure}

The shape of a cell in a valley in top view is depicted in Fig. \ref{schema_modele}(a).

In the following, we make the assumption that the sizes of a cell in the 3 directions are small as compared to the wavelength of the substrate profile.

The surface of the apical face is assumed to be flat and, given the condition of orthogonal sides, $S_{a}$ is related to $S_{\textsc{\tiny 1/2}}$ by:
\begin{equation}
S_{a} = S_{\textsc{\tiny 1/2}}(x)\left( 1 - \frac{1}{2}H(x)z''(x) \right)
\end{equation}

It explicitly depends on the curvature of the substrate.
On crests, where $z''$ is negative, the apical surface is larger than the surface at half height.
Conversely, in valleys, the apical surface is smaller than the surface at half height, as shown in the Fig. \ref{schema_modele}(a).

The basal surface $S_b$ is curved since it follows the substrate, and
\begin{equation}
S_{b} = S_{\textsc{\tiny 1/2}}(x)\left( 1 + \frac{1}{2}H(x)z''(x) \right) 
\end{equation}

This result for the basal surface $S_b$ is symmetrical to that obtained for the apical surface $S_a$ even though the two surfaces have very different geometries: curved for $S_b$ and flat for $S_a$.
The curvature of the basal surface has no influence on its area, up to first order in $\sqrt{S_{\textsc{\tiny 1/2}}}z''$.

The curvature of the substrate also has an influence on $S_{cc}$: 
\begin{equation}
S_{cc} = \alpha\sqrt{S_{\textsc{\tiny 1/2}}(x)}H(x)\left( 1 - \frac{\alpha ^2}{16\pi ^2}\frac{S_{\textsc{\tiny 1/2}}(x)}{H(x)} z''(x)\right)
\end{equation}

Finally, the expression of the cell volume is different from the case of a flat substrate.
The use of the expression $V = HS_{\textsc{\tiny 1/2}}(x)$ would lead to overestimating the volume of cells in the valleys and underestimating it on the crests (cf Fig. \ref{schema_modele}(b)).
The volume of a cell up to first order in $\sqrt{S_{\textsc{\tiny 1/2}}}z''$, reads:
\begin{equation} 
V=HS_{\textsc{\tiny 1/2}}(x) - \frac{\alpha ^4}{256 \pi ^3} S_{\textsc{\tiny 1/2}}^2 z''(x)
\label{equation_volume_courbe}
\end{equation}

\subsection{Energy of the cell}

Following previous works \cite{graner_equilibrium_2000} \cite{farhadifar_influence_2007} \cite{bi_motility-driven_2016} \cite{Hannezo2014}, we make the assumption that the equilibrium state of the epithelium is described by the minimization of an effective energy. 
In a simple model, we consider an average cell and minimize its effective $E_c$. 
We consider the following contributions to $E_c$: a apical energy per unit surface area (or surface tension) $\gamma_a$, mainly generated by the tension of the acto-myosin cortex of the apical face; a cell-substrate surface tension $\gamma_b$, which has mainly two contributions, a positive cortex tension and a negative cell-substrate adhesion energy; a cell-cell lateral surface tension $\gamma_{cc}$, with positive tensions from the cortex of the two cells and a negative cell-cell adhesion contribution; and finally an energy associated to the tension of the apical actomyosin belt, proportional to the apical perimeter $P_a$ and characterized by a line tension $\Lambda_a$:

\begin{equation}
E_{c} = \gamma_{a}S_{a} + \gamma_{b} S_{b} + \frac{\gamma_{cc}}{2} S_{cc} +\Lambda_{a}P_a
\label{energie_ligne}
\end{equation}

Surface and line tensions, $\gamma_{a}$, $\gamma_{b}$, $\gamma_{cc}$, and $\Lambda_a$ are considered uniform within a cell.
We now derive the energy of a cell on flat and sinusoidal substrates and minimize it to infer the equilibrium shape of the cell, \textit{i.e.} its height $H$ and surface at half height $S_{\textsc{\tiny 1/2}}$. 
All the energy minimizations will be performed under constant cell volume $V_0$, which is equivalent to adding a volume compressibility term $E_v = B (V - V_0)^2$ with a very large compression modulus B.

\subsubsection{Flat Substrate}
\label{energy_flat}

On a flat substrate, $S_a = S_b = S_{\textsc{\tiny 1/2}}$, (\textit{cf.} \ref{geometry_flat}), hence $V = HS_{\textsc{\tiny 1/2}}(x)$, $ P_a = \alpha \sqrt{S_{\textsc{\tiny 1/2}}}$, and the energy of the cell may be written as a function of $S_{\textsc{\tiny 1/2}}$ only:

\begin{equation}
E_{c} = \gamma_{a}S_{\textsc{\tiny 1/2}} + \gamma_{b} S_{\textsc{\tiny 1/2}} + \frac{\gamma_{cc}}{2} \alpha\frac{V}{\sqrt{S_{\textsc{\tiny 1/2}}}} + \Lambda_{a} \alpha \sqrt{S_{\textsc{\tiny 1/2}}}
\end{equation}

We introduce the dimensionless surface tension:
\begin{equation}
\gamma=\frac{\gamma_{a} + \gamma_{b}}{\gamma_{cc}}
\end{equation}
and the reduced line tension, $\Lambda$, which is a length:
\begin{equation}
\Lambda = \frac{\Lambda_a}{\gamma_{cc}}
\end{equation}

The reduced energy of the cell, which is a surface, then reads:
\begin{equation}
\frac{E_{c}}{\alpha \gamma_{cc}} = \frac{\gamma}{\alpha}S_{\textsc{\tiny 1/2}}  + \frac{V}{2\sqrt{S_{\textsc{\tiny 1/2}}}} + \Lambda \sqrt{S_{\textsc{\tiny 1/2}}}
\label{enegie_cellule}
\end{equation}

Its minimization at constant volume leads to the relation:
\begin{equation}
\gamma = \frac{\alpha}{4}\frac{1}{\sqrt{S_{\textsc{\tiny 1/2}}}} \left( H - 2\Lambda \right) 
\label{relation_lambda_gamma}
\end{equation}
The two parameters of the model are $\gamma$ and $\Lambda$, the other quantities, $\alpha$, $S_{\textsc{\tiny 1/2}}$ and $H$ are measured from experiments, leading to a quantitative relation between $\gamma$ and $\Lambda$.

\subsubsection{Curved substrate}

On a sinusoidal substrate, the reduced energy of the cell is obtained by using the expressions for the geometrical quantities derived in \ref{geometry_curved}:
\begin{align}
\begin{autobreak}
\frac{E_{c}}{\alpha \gamma_{cc}} =
- \frac{\alpha^2}{32\pi^2} \left(1 - \frac{\alpha^2}{16\pi} \right) z''(x)S_{\textsc{\tiny 1/2}}^{3/2}(x) 
+ \frac{\gamma}{\alpha}S_{\textsc{\tiny 1/2}}(x) 
+ \frac{V}{2\sqrt{S_{\textsc{\tiny 1/2}}(x)}} \left( 1 -  \frac{1}{2} \Lambda z''(x) \right) 
+ \Lambda \sqrt{S_{\textsc{\tiny 1/2}}(x)}
\end{autobreak}
\end{align}

The minimization of this reduced energy gives the preferred value of $S_{\textsc{\tiny 1/2}}$ for every value of the curvature $z''$. 
There is only one adjustable parameter since $V$ and $\alpha$ are measured on the samples and the measurements in the absence of curvature set a relation between $\gamma$ and  $\Lambda$, as discussed in \ref{energy_flat} (Equation \ref{relation_lambda_gamma}). 
In the following the adjustable parameter is $\Lambda$.

\section{Comparison with experimental measurements}
\label{section_comparison}
For cells cultured on flat substrates, we measured the following values of the geometrical parameters: mean height $H_0 = \SI{5.1 (1)}{\micro\meter}$, mean surface area $S_{\textsc{\tiny 1/2}} = \SI{80(7)}{\square\micro\meter}$, mean volume $V=H_0 \cdot S_{\textsc{\tiny 1/2}} = \SI{408}{\cubic\micro\meter}$, mean shape index $\alpha = 4.0 \pm 0.3$ (data are displayed as mean $\pm$ 95\%CI).

For each value of the curvature in the experimentally tested range, and for chosen values of the adjustable parameter $\Lambda$ in range [\SI{-100}{\micro\meter} ; \SI{100}{\micro\meter}], we find numerically the preferred value of $S_{\textsc{\tiny 1/2}}$ that minimizes $E_c$, deduce the preferred value of $H$ using equation \ref{equation_volume_courbe} and finally the relative height difference $\Delta h^i$ for two opposite values of the curvature.

\begin{figure}[h]
	\centering
	\includegraphics[width=0.8\columnwidth]{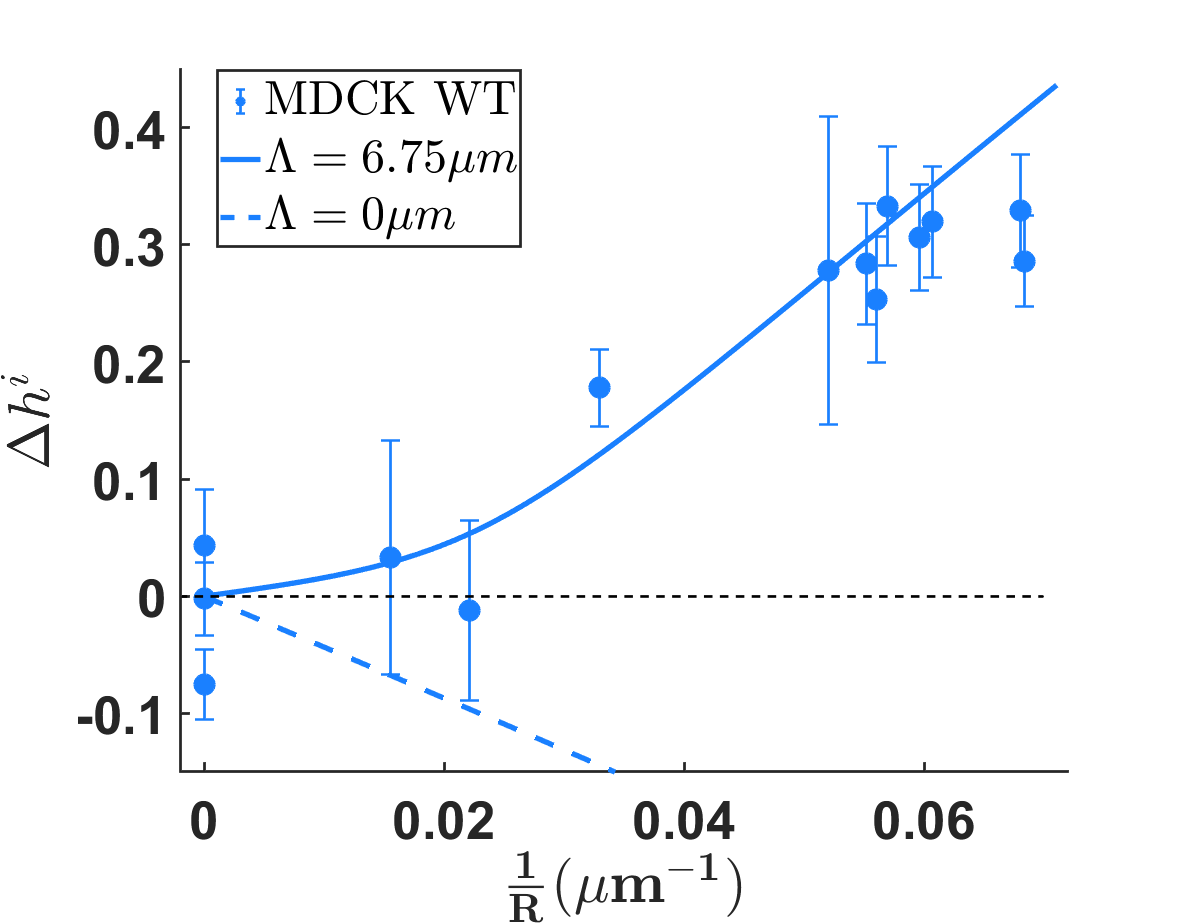}
	\caption{Relative height difference $\Delta h^i$ between cells on the crests and cells in the valleys as a function of $\frac{1}{R}$. The dotted curve corresponds to the model with only surface tensions. The solid curve corresponds to the best fit of the model with surface tensions and apical line tension.}
	\label{sinus_WT_0bleb_modele}
\end{figure}

In our experiments, we could test substrates with curvatures up to $\frac{1}{R}\approx \SI{0.068}{\per\micro\meter}$, which corresponds to a radius of curvature $R \approx \SI{15}{\micro\meter}$. 
The relative height difference $\Delta h^i$ is not significantly different from zero up to $\frac{1}{R}\approx \SI{0.03}{\per\micro\meter}$ and then increases with the curvature of the substrate, as reported in Figure \ref{sinus_WT_0bleb_modele}.

The dotted line in Figure \ref{sinus_WT_0bleb_modele} corresponds to the prediction of a model with cell surface energies only, obtained by setting  $\Lambda_a=0$ in the present model. 
This is the model that draws the most direct analogy between a cell and a liquid drop or a soap bubble. 
It has no adjustable parameter, since Equation \ref{relation_lambda_gamma} with $\Lambda=0$ sets the value of $\gamma$ from the measured volume, surface and shape index of cells on flat substrates. 
Such a model can clearly not account for our measurements since it predicts negative values of $\Delta h^i$, \textit{i.e.} cells thicker on crests than in valleys. 
The line tension is therefore the indispensable ingredient of the model in order to account for the measurements.

The best agreement with experimental results is obtained for $\Lambda=\SI{6.75}{\micro\meter}$ and is displayed in Figure \ref{sinus_WT_0bleb_modele} as a plain line. 

The apical line tension has been measured in the drosophila embryo by laser manipulation: $\Lambda_a \sim \SI{100}{\pico\newton}$ \cite{bambardekar_direct_2015}, which gives, with an intercellular tension of the order of $\gamma_{cc} \sim \SI{0.1}{\milli\newton\per\meter}$ \cite{salbreux_actin_2012}, $\Lambda \sim \SI{1}{\micro\meter}$.
The apical line tension in drosophila embryo has also been measured by laser ablation: $\Lambda_a \sim 1-10 \SI{}{\nano\newton}$ \cite{solon_pulsed_2009} which leads to $\Lambda \sim 10-100 \SI{}{\micro\meter}$.
Our measurement of $\Lambda$ is thus, within the same order of magnitude.

With $\Lambda = \SI{6.75}{\micro\meter}$, Equation \ref{relation_lambda_gamma} gives $\gamma = -0.94$. 
A negative value of $\gamma$ implies either $\gamma_{cc} <0$ or $\gamma_a + \gamma_b < 0$.
The possibility of a negative intercellular tension can be ruled out because all measurements have given positive values for this surface energy \cite{Maitre2012} \cite{solon_pulsed_2009}.
This leaves the possibility of $\gamma_a + \gamma_b < 0$ which implies $\gamma_b < 0$ since the apical surface energy $\gamma_a$ is necessarily positive.
A negative basal surface energy is possible since the adhesion of the cell with the substrate gives a negative contribution to $\gamma_b$. 
If the adhesion energy exceeds the positive contribution of the actomyosin contractility, then $\gamma_b$ is negative.

\section{Influence of biological parameters}
\label{section_biology}
\subsection{Intercellular adhesion protein E-cadherin}

We also used a genetically modified MDCK cell line that stably expresses the protein EcadGFP, the fusion of E-cadherin with GFP \cite{adams_mechanisms_1998}.
We expect over-expression of E-cadherin to have an impact in particular on the value of the intercellular surface energy, given the central role of E-cadherins in intercellular adhesion. 
The expression of EcadGFP is also of interest for imaging since it provides a well localized fluorescent signal at intercellular junctions, thus making the extraction of cell contours from microscopy images easier.

For EcadGFP MDCK cells, the measured area at mid-height $S_{\textsc{\tiny 1/2}}$ and shape index $\alpha$ in the absence of curvature have, within the margin of error, the same values as for WT MDCK cells with $S_{\textsc{\tiny 1/2}} = \SI{79(9)}{\square\micro\meter}$ and $\alpha = 4.0 \pm 0.3$. 
On the contrary, their mean height on flat substrates is different from that of WT MDCK cells. 
It is measured in the same way as for the WT MDCK cells, \textit{i.e.} as the average of the heights in the valleys and on the crests, and we obtain
\begin{equation}
H_0^{\text{\tiny{EcadGFP}}} = \SI{6.0(1)}{\micro\meter}
\end{equation}
As a result, the volume of the EcadGFP MDCK cells, $V = S_{\textsc{\tiny 1/2}} H = \SI{474(48)}{\cubic\micro\meter}$, is greater by $16 \%$ than the volume of WT MDCK cells. 

\begin{figure}[h]
	\centering
	\includegraphics[width=0.8\columnwidth]{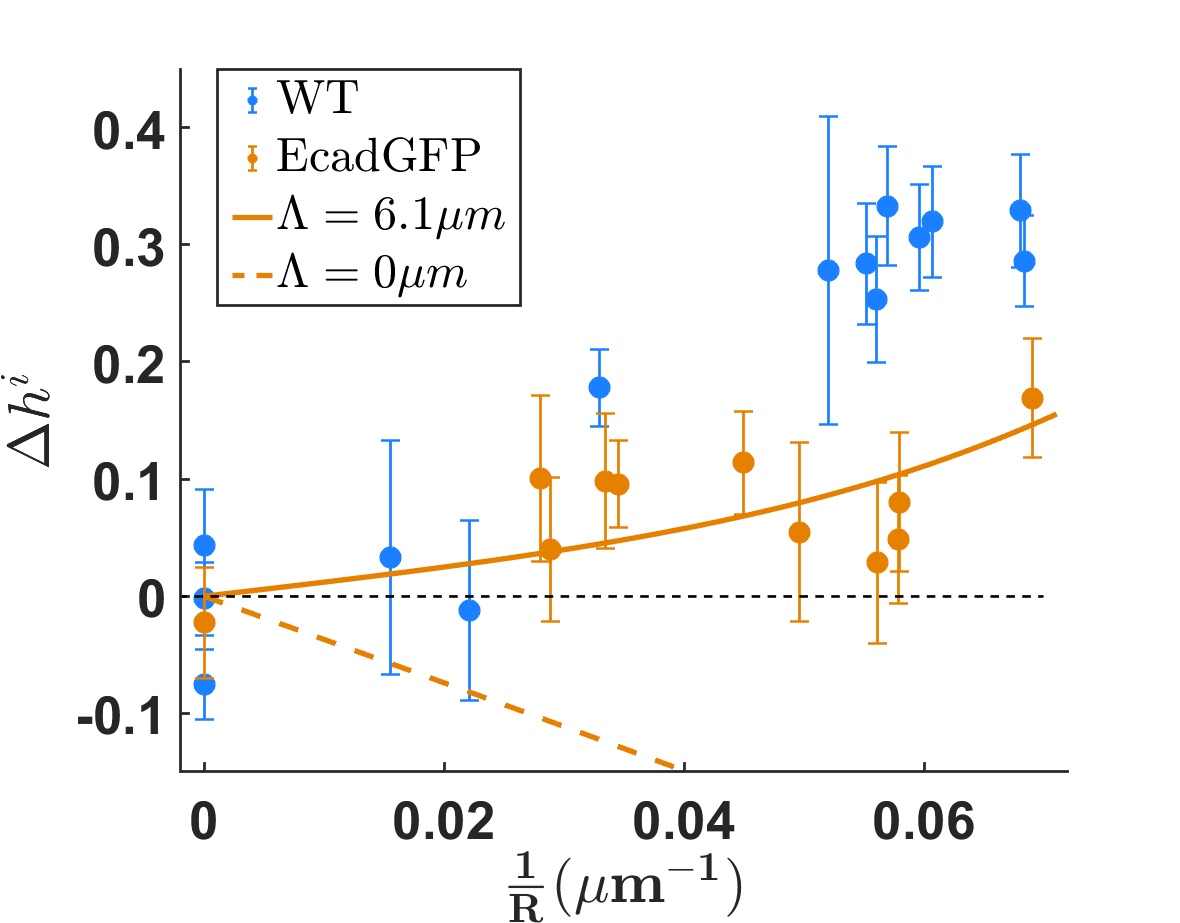}
	\caption{Relative height difference between cells on crests and cells in valleys $\Delta h^i$ as a function of $\frac{1}{R}$. The dotted curve corresponds to the model with only surface tensions. The solid curve corresponds to the best fit of the model with surface tensions and apical line tension.}
	\label{sinus_CadhGFP_0bleb}
\end{figure}

The relative height difference $\Delta h^i$ on sinusoidally-shaped substrates as a function of curvature is displayed in Figure \ref{sinus_CadhGFP_0bleb}, along with the measurements on WT cells. 
The heights of 1310 EcadGFP cells were measured in 11 samples from 6 independent experiments.

The height of the EcadGFP MDCK cells is significantly less sensitive to curvature than that of WT MDCK cells. 
For the intermediate values of the curvature, $1/R \sim \SI{0.03}{\per\micro\meter}$, the relative height difference is smaller for EcadGFP MDCK cells ($\Delta h^i = 0.10 \pm 0.05$) than for WT MDCK cells ($\Delta h^i = 0.18 \pm 0.03$). 
The difference between the two cell lines is statistically significant for all samples with curvature larger than \SI{0.04}{\per\micro\meter}, with $\Delta h^i$ always smaller than $0.2$ for EcadGFP MDCK cells.
As for the WT MDCK cells, the simplest model with surface energies only, displayed as a dotted line in Figure \ref{sinus_CadhGFP_0bleb}, cannot account for the measurements.
The best agreement between the experimental data and the complete model, with apical line tension, is obtained for $\Lambda = \SI{6.1}{\micro\meter}$.
The value of $\Lambda = \frac{\Lambda_a}{\gamma_{cc}}$ is therefore smaller for EcadGFP MDCK cells than for WT MDCK cells.
This implies a lower apical line tension for EcadGFP MDCK cells than for WT MDCK cells or a higher intercellular surface energy.
EcadGFP MDCK cells should express more E-cadherins than the WT line because they express EcadGFP in addition to E-cadherins.
One expects this overexpression of intercellular adhesion proteins to give a lower intercellular surface energy $\gamma_{cc}$.
The effect on apical line tension is less obvious.
However, we note that the difference in reduced apical line tension $\Lambda$ between the WT MDCK cells and the EcadGFP MDCK cells is very small ($\sim$10\%).

With $\Lambda = \SI{6.1}{\micro\meter}$ using equation \ref{relation_lambda_gamma}, we get $\gamma = -0.69$.
Thus, as for the WT MDCK cells, we obtain a negative value for $\gamma$, which implies $\gamma_a + \gamma_b < 0$ and therefore $\gamma_b < 0$.
As already discussed in section \ref{section_comparison}, this implies that the negative contribution of cell-substrate adhesion to the surface energy exceeds the positive contribution of actomyosin contractility.
In addition, the dimensionless surface tension $\gamma$ is smaller, in absolute value, for EcadGFP MDCK cells than for the WT MDCK cells ($\gamma^{EcadGFP} = -0.69$ and $\gamma^{WT} = -0.94$, respectively).
\begin{equation}
\left| \frac{\gamma_{a} + \gamma_{b}}{\gamma_{cc}}\right|^{WT} > \left| \frac{\gamma_{a} + \gamma_{b}}{\gamma_{cc}}\right|^{EcadGFP}
\label{equation_comparaison_gamma}
\end{equation}

Equation \ref{equation_comparaison_gamma} can be satisfied if the apical surface tension of EcadGFP MDCK cells is greater than the apical surface tension of WT MDCK cells.
Increased apical surface tension could be the result of E-cadherin overexpression, leading to more mechanical coupling between the cortices of neighbouring cells, as was observed in previous studies \cite{Foty2005a}  \cite{maitre_adhesion_2012}.

\subsection{Myosin-II activity}
\label{myosin_activity}

We also altered the activity of myosin-II, which triggers the contractility of the cortex \cite{Murrell2015} and also plays an important role in the tension exerted by the actin belt that connects the adherens junctions \cite{harris_adherens_2010}.
The tension of the actomyosin cortex and of the actin belt are expected to scale with myosin-II activity.
Myosin-II activity can be inhibited with blebbistatin. We used two concentrations of blebbistatin, $\SI{50}{\micro M}$, which saturates the decrease in myosin-II contractility, and $\SI{5}{\micro M}$, which gives an intermediate effect at around $50\%$ decrease of myosin-II activity \cite{mitrossilis_single-cell_2009}.

First, the geometric parameters of the cells were measured in the absence of curvature. 
The addition of $\SI{5}{\micro M}$ blebbistatin increased the mean height of WT MDCK cells on flat substrates and decreased that of EcadGFP MDCK cells: $H_0^{\SI{5}{\micro\Molar}bleb} = \SI{5.6(1)}{\micro\meter}$ , for both cell types (cf. Table \ref{table_mesure_parametres}). 
The measured shape index and cell surface had within the margin of error the same values as without blebbistatin: $\alpha = 4.0 \pm 0.4$, $S_{\textsc{\tiny 1/2}}^{\SI{5}{\micro\Molar}bleb} = \SI{81(8)}{\square\micro\meter}$ for WT MDCK cells and $77 \pm 10 \mu m^2$ for EcadGFP MDCK cells (cf. Table \ref{table_mesure_parametres}).

\begin{figure}[h]
	\begin{center}
		\includegraphics[width=0.9\columnwidth]{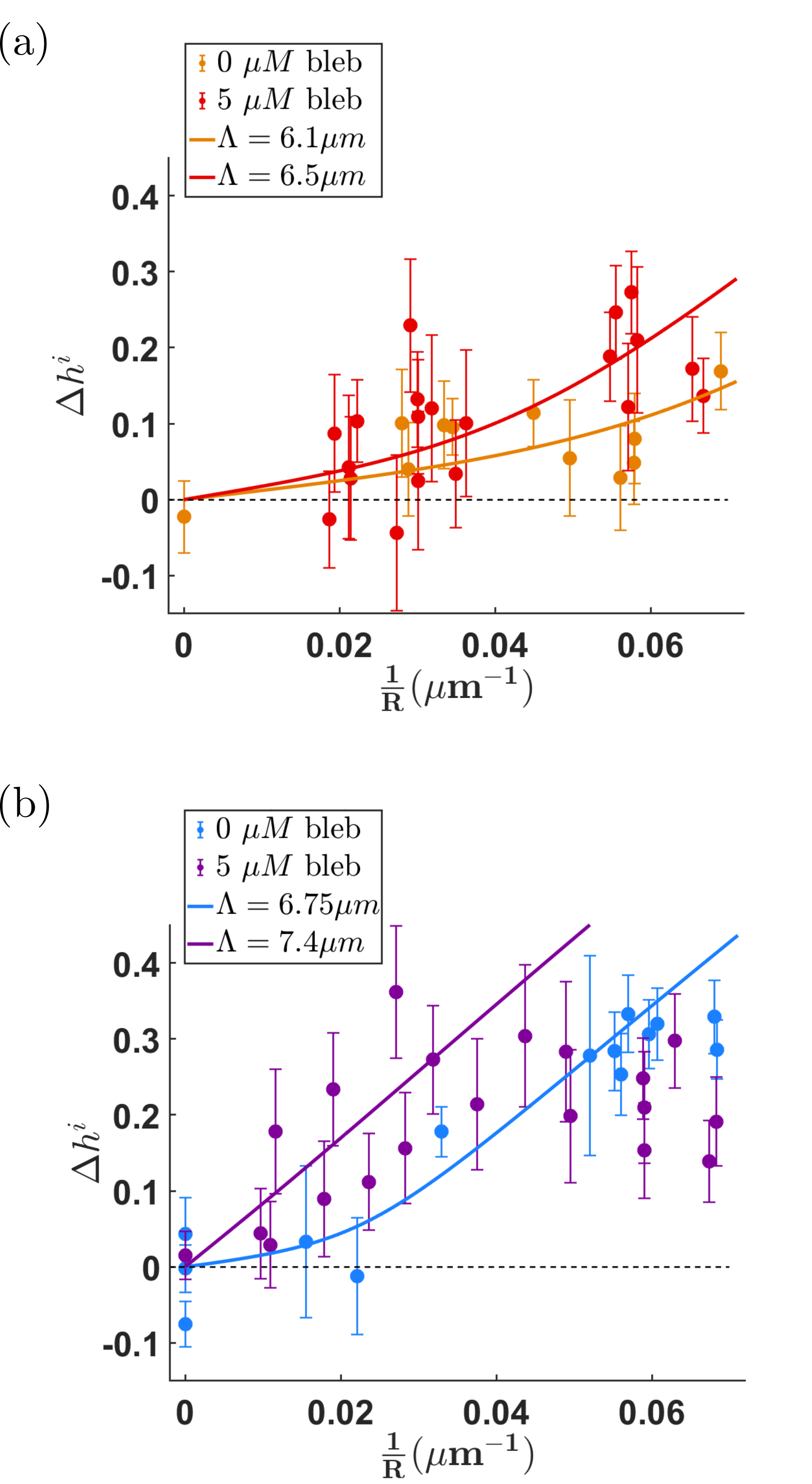}
	\end{center}
	\caption{Relative height difference between cells on crests and cells in valleys $\Delta h^i$ as a function of $\frac{1}{R}$. The solid curves correspond to the best fits of the model with surface tensions and apical line tension. (a) EcadGFP MDCK cells. (b) WT MDCK cells.}
        \label{sinus_bleb}
\end{figure}

The influence of $\SI{5}{\micro\Molar}$ blebbistatin on the height of MDCK cells on curved substrates is displayed in Fig.\ref{sinus_bleb}(a) and Fig. \ref{sinus_bleb}(b) for EcadGFP and WT cells, respectively. 
The height of 7430 cells was measured in 6 separate experiments, including 3 experiments on samples treated with blebbistatin. 
For EcadGFP MDCK cells, the relative height difference $\Delta h^i$ between positively and negatively curved substrates was globally larger with than without blebbistatin, but the difference was not statistically significant except for the measurements around $\frac{1}{R} \approx \SI{0.057}{\per\micro\meter}$. 
For WT MDCK cells, the addition of $\SI{5}{\micro\Molar}$ blebbistatin altered the dependence of $\Delta h^i$ on curvature. 
For low and intermediate curvature $\left( \frac{1}{R} \leq \SI{0.04}{\per\micro\meter}\right)$, the addition of blebbistatin increased the relative height difference between valleys and crests. 
On the contrary, for larger curvature, the relative height difference after addition of blebbistatin saturated at $\Delta h^i \approx 0.30$ and even decreased down to about 0.2 for the largest curvatures $\left( \frac{1}{R} \approx \SI{0.068}{\per\micro\meter} \right)$. 

The relative dispersion of experimental data was larger in experiments with blebbistatin, both for WT MDCK cells and EcadGFP MDCK cells.
This could be due to the fact that the duration of blebbistatin treatment was not long enough to allow the cells to fully change their shape accordingly.
Multiple time scales are involved in such processes\cite{wyatt_question_2016}, presumably up to tens of minutes, but the treatment duration of 15 minutes for these experiments was chosen in order to preserve the structural integrity of the cells\cite{liu_mechanical_2010}.

As for the other measurements, a model with surface energies only cannot account for the measurements of $\Delta h^i$ with blebbistatin.

The model using surface energies and apical line tension accounts for the measurements on EcadGFP MDCK cells and for the measurements on WT MDCK cells at low and intermediate curvature. 
The best adjustments of the experimental points, restricted to $\frac{1}{R}<\SI{0.04}{\per\micro\meter}$ for WT MDCK cells, are obtained for $\Lambda_{EcadGFP}^{\SI{5}{\micro\Molar} bleb} = \SI{6.5}{\micro \meter}$ and $\Lambda_{WT}^{ \SI{5}{\micro\Molar} bleb}=7.4 \mu m$. 
They are displayed as a red line in Fig. \ref{sinus_bleb}(a) and as a purple line in Fig. \ref{sinus_bleb}(b), respectively. 
Using Eq. \ref{relation_lambda_gamma} we obtain $\gamma_{EcadGFP}^{\SI{5}{\micro\Molar} bleb}=-0.83$ and $\gamma_{WT}^{\SI{5}{\micro\Molar} bleb}=-1.03$.

\begin{table}
\caption{\label{table_mesure_parametres} Measurements of the shape ($H_0$ and $S_{\textsc{\tiny 1/2}}$) of MDCK cells and parameters of the model that fit the experimental data ($\Lambda$ and $\gamma$). Note that only $\Lambda$ is a free parameter of the model.}
\begin{ruledtabular}
\begin{tabular}{lcccc}
 &\multicolumn{2}{c}{WT}&\multicolumn{2}{c}{EcadGFP}\\
 &\SI{0}{\micro\Molar} bleb&\SI{5}{\micro\Molar} bleb&\SI{0}{\micro\Molar} bleb&\SI{5}{\micro\Molar} bleb\\
\hline
$H_0 (\SI{}{\micro\meter})$ & 5.1 & 5.6 & 6.0 & 5.6\\
$S_{\textsc{\tiny 1/2}} (\SI{}{\square\micro\meter})$ & 80 & 81 & 79 & 77\\
$\Lambda (\SI{}{\micro\meter}) $ & 6.75 & 7.4 & 6.1 & 6.5 \\
$\gamma$ & -0.94 & -1.03  & -0.69  & -0.83 \\
\end{tabular}
\end{ruledtabular}
\end{table}

The addition of \SI{5}{\micro\Molar} blebbistatin reduces the value of $\gamma$ (increases its absolute value), by about 10\% for WT MDCK cells and by about 20\% for EcadGFP MDCK cells, and increased the value of $\Lambda$, by about 10\% for WT MDCK cells and by about 7\% for EcadGFP MDCK cells.

It is expected that the alteration in myosin-II activity will influence the parameters of the model : $\gamma$ and $\Lambda$ \cite{furukawa_epithelial_2017} \cite{uyeda_stretching_2011}.
Indeed, decreasing the activity of myosin-II by adding blebbistatin reduces the contractility of the actomyosin cortex.
This can result in a decrease in all the surface energies involved in the model.
The activity of myosin-II is also expected to impair cell-substrate and cell-cell adhesion, which should increase the values of $\gamma_b$ and $\gamma_{cc}$ by decreasing adhesion energies.
Indeed we observed that when increasing the concentration of blebbistatin to \SI{50}{\micro\Molar}, the epithelium detached from the substrate (see Appendix \ref{detachement} for more detail), making measurements impossible, and supporting the idea that blebbistatin decreases cell adhesion to the substrate.
Thus, blebbistatin should mainly decrease apical surface tension $\gamma_a$ compared to other surface tensions.
The net effect on the value of the reduced surface tension $\gamma$ is therefore difficult to predict. 
Myosin-II activity also influences the contractility of the actin belt that connects the adherent junctions, which should decrease the apical line tension $\Lambda_a$.
The increase that we observed in the value of $\Lambda$ suggests a larger relative decrease in the value of $\gamma_{cc}$ than in the value of $\Lambda_a$ when adding blebbistatin.

Table I summarizes the parameters measured in the absence of curvature ($H_0$ and $S_{\textsc{\tiny 1/2}}$) for each condition, along with the parameters obtained by fitting the experimental measurements on curved substrates by the model ($\Lambda$ and $\gamma$).

We found that the relative surface tension $\gamma=\frac{\gamma_a+\gamma_b}{\gamma_{cc}}$ is always negative and ranges from $-1.03$ and $-0.69$. 
As already discussed in section \ref{section_comparison}, this shows that the tension of the basal surface $\gamma_b$ is negative and that the different surface energies are all of the same order of magnitude. 

The value of $\Lambda$ varied by less than 10\% around a mean value $\Lambda \approx \SI{6.7}{\micro\meter}$, regardless of E-cadherin overexpression or addition of blebbistatin. 
This suggests a regulation of this parameter. 
Indeed $\Lambda_a$ and $\gamma_{cc}$ are both related to properties of cell-cell contact zone, in particular via E-cadherins which are involved in the structure of both cell-cell adhesions and adherens junctions.

For WT MDCK cells with \SI{5}{\micro\Molar}, the model could not account for the saturation of the relative height difference at $\Delta h^i \approx 0.30$ observed in the experiments for high curvatures ($\frac{1}{R} > \SI{0.04}{\per\micro\meter}$).
A closer look to the measurements (Fig. \ref{sinus_bleb}(a) and \ref{sinus_bleb}(b) suggests that such a saturation may also be observed for WT MDCK cells without blebbistatin at $\Delta h^i \approx 0.30$ when $\frac{1}{R} > \SI{0.055}{\per\micro\meter}$, and for EcadGFP MDCK cells with \SI{5}{\micro\Molar} blebbistatin  at $\Delta h^i \approx 0.20$ when $\frac{1}{R} > \SI{0.058}{\per\micro\meter}$.
Our model could fail to account for the results at high curvature because its geometrical approximations are not fullfilled anylonger. 
Yet since the cells have about the same sizes in all four conditions, saturation should happen for the same values of $\frac{1}{R}$ and/or $\Delta h^i$ on all four curves.
The saturation could also come from a missing ingredient in the energy of the cell that would tend to resist to excessive cell flattening.
Modeling the resistance to compression of the cell contents as that of gaussian biopolymers, as previously proposed \cite{Hannezo2014}, failed to account quantitatively for our results. 

\section{Conclusions and outlook}
\label{conclusion}

Using substrates with a sinusoidal profile, we explored the role of curvature in the 3D shape of epithelial cells. 
We demonstrated that for large curvature of the substrate (curvature radius smaller than about \SI{40}{\micro\meter}), cells are thicker on positive curvatures of the substrate (in valleys) than on negative curvatures (on crests), with a relative thickness difference increasing with the curvature. To account for these measurements, we developed a simple 3D energetic cellular model, inspired by existing vertex models. Since such models often come with many parameters, whose values have to be adjusted, we focused on developing a minimal model pointing on the key biological ingredients that shape epithelial cells. We showed that a minimal model with surface tensions and an apical line tension as the physical ingredients for the energy accounts for the shape of epithelial cells on curved substrates, the apical line tension being a necessary ingredient of the model. This minimal model, which has only one adjustable parameter, accounted for measurements in a wide range of experimental conditions, with WT MDCK cells as well as cells over-expressing E-cadherin, and with cells with a decreased contractility. Nevertheless, for the highest curvatures and for cells with inhibited myosin-II activity, the effect of curvature on cell thickness saturated, pointing at the need to add in the model a term limiting large cell deformation in extreme cases. The experiments we set up in this study could be used to test other lines of epithelial cells, in particular ones with different morphologies, for instance columnar cells like Caco-2, since their surface tensions and line tension parameters should be very different. In the future, our simple model will also be used to get insight in the cross-talk between curvature and cell fate in tissues such as the intestinal microvilli. 

\appendix
\section{Materials and methods}
\label{Materials and methods}

\subsection{Cell culture}

MDCK-II cells from \textit{European Collection of Authenticated Cell Cultures} (ECACC) and MDCK-II cells genetically modified to stably express E-cadherin-GFP\cite{adams_mechanisms_1998} were grown in \textit{Dulbecco’s Modifed Eagle Medium} (DMEM), supplemented with 10\% fetal bovine serum (FBS), 100U/mL \textit{penicillin-streptomycin}, and \SI{50}{\micro\gram\per\milli\liter} \textit{G418} (Geneticin) for EcadGFP MDCK cells.

For the experiments, cells were trypsinised from culture flasks and seeded on flat or sinusoidal substrates, at a density of 1900 cells/\SI{}{\square\milli\meter} allowing  confluence to be reached in 48h.
The medium was renewed 48h after seeding.
For some of the experiments, blebbistatin was added to the culture medium at a final concentration of \SI{5}{\micro\Molar} or \SI{50}{\micro\Molar} for 15 minutes before fixing the cells.

\subsection{Immunofluorescence}

\subsubsection{Fixation}

The cells were fixed with Paraformaldehyde (PFA) 4\% in phosphate-buffered saline (PBS) 96h after seeding.

\subsubsection{Staining}

The nucleus was labeled with DAPI, \SI{0.6}{\micro \Molar} in PBS, for 20 min at \SI{4}{\celsius}.
F-actin was labeled with \textit{SiR-actin} (Spirochrome), \SI{100}{\nano\Molar} in PBS for 12h at \SI{4}{\celsius}.
The apical surface of the cells was labeled with mouse anti-gp135 (DSHB, reference 3F2/D8)\cite{herzlinger_mdck_1982} 1/25 for 20 min at \SI{4}{\celsius} and goat anti-mouse Dylight 549 (Abcam) for 30 min at \SI{4}{\celsius}.

The substrates were stained with labeled fibronectin (DyLight fast conjugaison kit, Abcam). 
We used DyLight 488 for WT MDCK cells and DyLight 549 for EcadGFP MDCK cells.

The different fluorophores used for each MDCK cell line are summarized in table \ref{tableau_fluo}.
\begin{table}
\caption{\label{tableau_fluo} Fluorophores used to label the contours of the cells in 3D (gp135, F-actin, E-cadherin, fibronectin) and their positions (nucleus).}
\begin{ruledtabular}
\begin{tabular}{lcc}
     & Wild type & EcadGFP \\ \hline
    gp135 (apical face)	& Dylight 549 	& Dylight 549 \\
    F-actin 			& SiR-actin 	& SiR-actin \\
    E-cadherin 			& none			& GFP \\
    Nucleus 			& DAPI 			& DAPI \\
    Fibronectin 		& Dylight 488  	& Dylight 549 \\
\end{tabular}
\end{ruledtabular}
\end{table}

\subsubsection{Imaging}

Image acquisition was performed using Metamorph and a motorized inverted microscope (Leica DMI8) equipped with a 63X water immersion objective, a wide field spinning disk head CSU-W1 (Yokogawa - Andor), and a sCMOS Orca-Flash 4 V2+ camera (Hamamatsu) resulting in a field of view of $\SI{230}{\micro \meter}\times \SI{230}{\micro \meter}$.

The sampling in the direction of the optical axis was \SI{0.25}{\micro \meter}.

\subsection{Image analysis}
\label{analyse_images}

MATLAB, ImageJ, and MIJ \cite{sage_mij_2012}, a tool for calling ImageJ instructions in MATLAB, were used for image analysis. 

The geometry of the sinusoidal substrate makes 3D image analysis difficult.
As a result, semi-automated measurements of cell height were limited to the crests and the valleys of the substrate.

\begin{figure}[h]
	\begin{center}        
		\includegraphics[width=0.9\columnwidth]{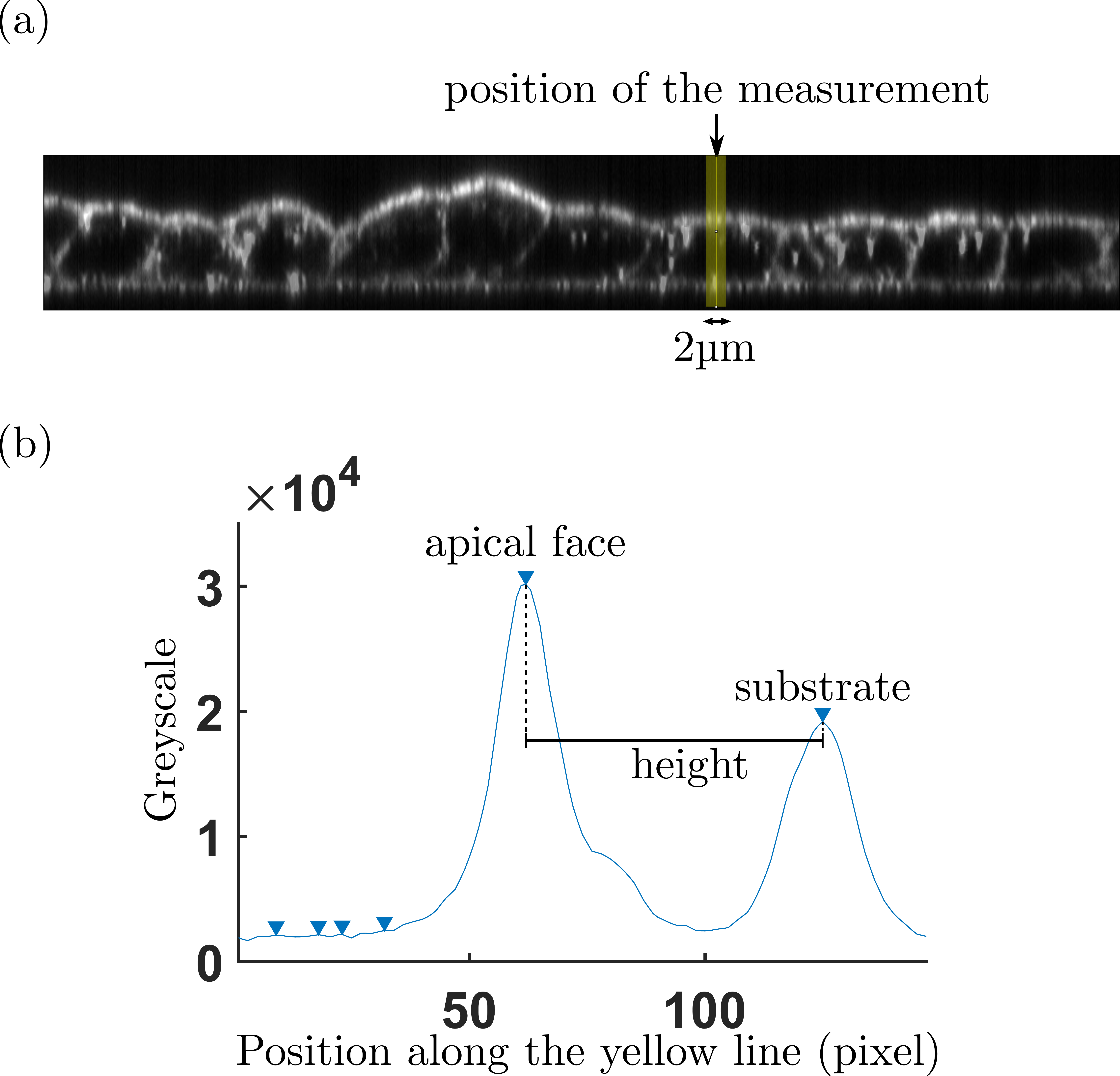}
	\end{center}
	\caption{Measurement of the height of the cells on a sinusoidal profile substrate. (a) Cross-section view along a valley, sum of the renormalized signals of gp135, F-actin and fibronectin. The yellow line gives the position of the profile measurement. The width of the measurement is \SI{2}{\micro\meter}. (b) Intensity profile along the yellow line. Local extrema are represented by a blue chevron. The two highest local extrema are considered to be the positions of the apical face and the substrate.}
        \label{Mesure hauteur sinus}
\end{figure}

First the user draws lines along each valley and each crest.
The program then automatically creates cross-sectional views along these lines.
Next, the positions of the cells on the cross-sectional view are detected using a threshold (Otsu method) on the DAPI signal.
The gp135, F-actin and fibronectin channels are summed-up, after intensity renormalization, to obtain an image that clearly displays the apical and basal faces of the cells.
The height of the cells are finally automatically measured.
A typical example is given in Fig. \ref{Mesure hauteur sinus}(a).
At each cell position (the center of the nucleus identified in the DAPI channel), the program plots the intensity profile along the apico-basal axis ($z$) averaged over a width of \SI{2}{\micro\meter} in the $y$ direction.
On this profile, the program identifies the two intensity local maxima that most likely correspond to the apical and basal position of the cell, respectively, as shown in the Fig. \ref{Mesure hauteur sinus}(b).
Each measurement is checked using criteria related to the position of the substrate, which cannot vary by more than \SI{1}{\micro\meter} between two neighbouring cells, and to the minimum thickness of a cell, set here at \SI{1}{\micro\meter}. 
About 5\% of the measurements are readjusted by hand and 1\% are discarded.

The last step consists of measuring the geometry of the substrate. 
The wavelength and amplitude of the sinusoidal profile are measured by hand from the positions of crests and valleys on a cross-sectional view orthogonal to the direction of the grooves in the substrate. 
However, the substrate profile sometimes differs from a perfect sinusoid (cf. Fig. \ref{fig_images_exp}(c). 
To take this asymmetry into account, an apparent wavelength and an apparent amplitude are also measured on the cross-section view as twice the distances between two successive inflection points on either side of a valley, in the $x$ and $z$ directions, respectively. 
The same applies to the crests. 
The respective curvatures of the valley and the crests are calculated from these apparent wavelengths and amplitudes, and the curvature of the substrate is taken as the average of the two curvatures. 
In order to avoid a systematic bias in the curvatures of the crests and valleys, experiments were performed on all positive and negative replicates of the substrates and no significant difference in the results was observed.

\subsection{Sinusoidally-shaped substrates}
\label{methods_substrate_making}

\begin{figure}[h]
	\centering
	\includegraphics[width=\columnwidth]{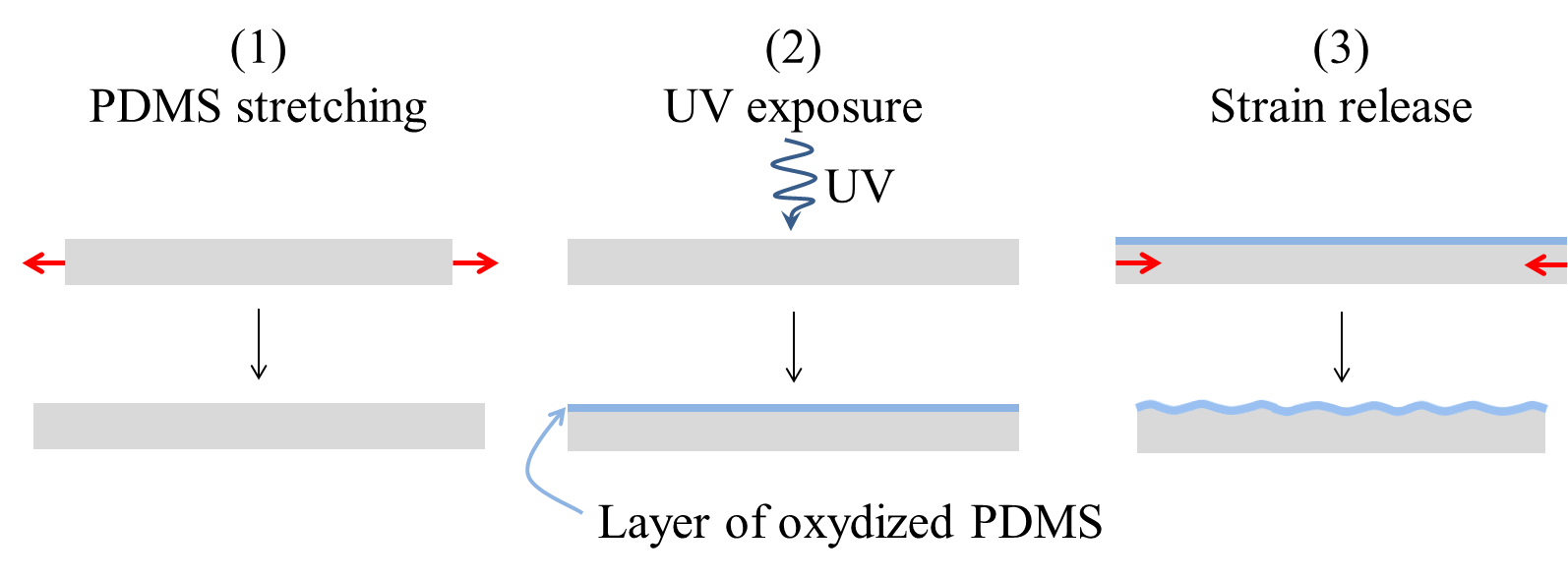}
	\caption{Fabrication of PDMS moulds with sinusoidal profile. The first step is to stretch a sample, the red arrows represent the applied strain. In (2) the sample is exposed to UV which forms a thin layer of oxidized PDMS which is stiffer than the rest of the sample. In (3) the stretch imposed in (1) is released and a sinusoidal surface profile is formed due to mechanical instability.}
	\label{schema_fabrication_moules}
\end{figure}

The creation of a sinusoidal height profile, which is not directly possible with conventional photolithography techniques, was achieved by exploiting a mechanical instability that creates sinusoidal patterns \cite{bowden_spontaneous_1998} \cite{brau_multiple-length-scale_2011}.
The steps of the manufacture of sinusoidal profile moulds are summarised in Figure \ref{schema_fabrication_moules}.
The principle is to stretch a sample of PDMS, then expose it to UV light and finally release the deformation.
UV exposure creates a thin layer of oxidized PDMS, which is more rigid.
When the deformation is released, wrinkles form to accommodate the disparity between the Young's moduli of the oxidized layer $E_s$ and of the rest of the sample $E_c$.
Both wavelength $\lambda$ and amplitude $A$ of the sinusoidal profile depend on the applied strain $\epsilon$ and on the thickness $h$ of the oxidized layer, which is varied by changing the duration of UV exposure\cite{brau_multiple-length-scale_2011}
\begin{equation}
\lambda \propto h \sqrt[3]{\frac{E_s}{E_c}}
\end{equation}
\begin{equation}
A \propto \lambda \sqrt{\epsilon}
\end{equation}

To obtain sinusoidal PDMS substrates from these moulds, they are first silanized with 1,1,2,2,-tridecafluoro-1,1,2,-tetrahydrooctyl-1-trichlorosilane in vapor phase to facilitate the release of the elastomer after curing ; PDMS (Sylgard 184 + 10\% cross-linker, DowCorning) is then poured over the silicon template, spin-coated for 90s at 450rpm to a thickness of $~\SI{200}{\micro\meter}$, cured overnight at \SI{65}{\celsius}, and finally peeled off.

Flat PDMS substrates are produced by a similar procedure using a simple flat epoxy disc as a mould.

\section{Derivation of the energy of the cell on curved substrate}

\subsection{Cell surface and volume}

This appendix gives a detailed derivation of the energy of the cell as written in Eq. \ref{enegie_cellule}.
The characteristic size of the cell in the median plane is defined as $R_{\textsc{\tiny 1/2}}=\frac{P_{\textsc{\tiny 1/2}}}{2\pi}$, where $P_{\textsc{\tiny 1/2}}$ is the cell perimeter in the median plane.
Since the intercellular junctions are assumed to be orthogonal to the substrate plane, the characteristic size of the apical surface modified by the curvature and in the orthogonal direction to the grooves is given by
\begin{align*}
\begin{autobreak}
2R_{a} 
= 2R_{\textsc{\tiny 1/2}}(x) 
+ \frac{1}{2}H(x)\left[ \frac{dz}{dx}(x-R_{\textsc{\tiny 1/2}}(x))-\frac{dz}{dx}(x+R_{\textsc{\tiny 1/2}}(x))\right] 
\end{autobreak}
\end{align*}

Considering that $R_{\textsc{\tiny 1/2}} \ll 6\frac{\lambda}{2\pi} $ we get 
\begin{equation*}
R_{a} = R_{\textsc{\tiny 1/2}}\left( 1 - \frac{1}{2}Hz'' \right) 
\end{equation*} 
hence the surface of the apical face 
\begin{equation*}
S_{a} = S_{\textsc{\tiny 1/2}}\left( 1 - \frac{1}{2}Hz'' \right) 
\end{equation*}

This expression for the apical surface $S_a$ gives an explicit dependence on the curvature $z''$ of the substrate.
On crests, where $z''$ is negative, the apical surface is larger than the surface at half height.
Conversely, in the valleys, the apical surface is smaller than the surface at half height, as shown in the Fig. \ref{schema_modele}(a).

The basal surface $S_b$ is curved since it follows the substrate. 
The height difference, in a cross-section, between the positions of the substrate at the edge of the cell and at its centre is denoted $\delta$, as depicted in Fig. \ref{schema_modele}(b). 
The value of $\delta$ depends on the direction $\theta$ of the cross-section with respect to the direction $x$ (as defined in Fig. \ref{schema_modele}).

By approximating the substrate profile by a parabola one obtains 

\begin{align*}
\begin{autobreak}
S_{b} = 
S_{\textsc{\tiny 1/2}} 
+ \int_{-\frac{\pi}{2}}^{\frac{\pi}{2}} R_{\textsc{\tiny 1/2}}^2 \cos^2\theta \left( \frac{H}{2} - \delta(\theta) \right)
\times \left( 2 + \left( \frac{H}{2} - \delta(\theta) \right) z'' \right) z'' \diff\theta 
+ \delta^2(\theta) \diff\theta 
\end{autobreak}
\end{align*}

Since $\delta(\theta) = z(x+R_{\textsc{\tiny 1/2}}\cos\theta) - z(x) - R_{\textsc{\tiny 1/2}}\cos\theta\frac{dz}{dx}$, assuming  that $R_{\textsc{\tiny 1/2}}\ll3\frac{\lambda}{2\pi}$, $\delta(\theta) = \frac{1}{2} (R_{\textsc{\tiny 1/2}}\cos\theta)^{2}z''$, which gives 

\begin{align*}
\begin{autobreak}
S_{b} = 
S_{\textsc{\tiny 1/2}}\left[1+ \frac{1}{2}Hz'' + \left( \frac{1}{8}H^2 - \frac{21}{64}R_{\textsc{\tiny 1/2}}^2 \right) \left(z''\right)^2\right]
+ S_{\textsc{\tiny 1/2}}\left[-\frac{3}{16}HR_{\textsc{\tiny 1/2}}^2\left(z''\right)^3 + \frac{5}{64}R_{\textsc{\tiny 1/2}}^4\left(z''\right)^4 \right] 
\end{autobreak}
\end{align*}

With $ R_{\textsc{\tiny 1/2}} \ll \frac{\lambda^2}{4A} $ 
\begin{equation*}
S_{b} = S_{\textsc{\tiny 1/2}}\left( 1 + \frac{1}{2}Hz'' \right) 
\end{equation*}

This expression for the basal surface $S_b$ is remarkably symmetrical to that obtained for the apical surface $S_a$ while these two surfaces have very different geometries: curved for $S_b$ and flat for $S_a$.
The curvature of the basal surface therefore has no influence on its expression, at the first order in $Rz''$.

To compute the lateral cell surface $S_{cc}$ the perimeter of the cell as a function of the position along the apico-basal axis has to be written.

At half height $P_{\textsc{\tiny 1/2}} = 2\pi R_{\textsc{\tiny 1/2}}$.
In the other positions, the expression is different because the cell surface is anisotropically deformed.
The characteristic size of the cell is still $R_{\textsc{\tiny 1/2}}$ in the $y$ direction but is $R(z) \approx R_{\textsc{\tiny 1/2}} (1-zz'')$ is the $x$ direction. 
By analogy with the approximate perimeter of an ellipse $\pi (a+b)$ with a and b the semi-major and semi-minor axes of the ellipse, the approximate perimeter of the cell at position $z$ is $P(z) \approx \pi (R_{\textsc{\tiny 1/2}} + R(z))$.

The intercellular contact surface is therefore given by 
\begin{align*}
\begin{autobreak}
S_{cc}
=\int_{-\frac{\pi}{2}}^{\frac{\pi}{2}} \diff\theta \int_{-\frac{H}{2}+\delta(\theta)}^{\frac{H}{2}} \left[ R_{\textsc{\tiny 1/2}} + R_{\textsc{\tiny 1/2}}\times \left( 1 - z z'' \right) \right] \diff z 
= \int_{-\frac{\pi}{2}}^{\frac{\pi}{2}} 2R_{\textsc{\tiny 1/2}}\left( H-\delta(\theta) \right) \left( 1 - \frac{1}{2}\delta(\theta)z'' \right) \diff\theta
= \alpha\sqrt{S_{\textsc{\tiny 1/2}}}H
\times \left( 1 - \frac{1}{4}\frac{R_{\textsc{\tiny 1/2}}^2}{H} z'' - \frac{1}{16} R_{\textsc{\tiny 1/2}}^2 \left( z'' \right)^2 + \frac{3}{64} \frac{R_{\textsc{\tiny 1/2}}^4}{H} \left( z'' \right)^3 \right)
\end{autobreak}
\end{align*}

Assuming that $ R_{\textsc{\tiny 1/2}} \ll \frac{\lambda^2}{4A} $ 
\begin{equation*}
S_{cc} = \alpha\sqrt{S_{\textsc{\tiny 1/2}}}H\left( 1 - \frac{1}{4}\frac{R_{\textsc{\tiny 1/2}}^2}{H} z''\right)
\end{equation*}

This expression shows that the curvature of the substrate has an influence on $S_{cc}$. 

The expression of cell volume also differs from the case of a flat substrate.
Indeed, the expression $V = HS_{\textsc{\tiny 1/2}}$ overestimates the volume of cells in the valleys because it also includes some volume of the substrate.
Similarly, this expression underestimates the volume on the crests.
The volume is therefore written as $V = HS_{\textsc{\tiny 1/2}} - \Delta V$ with $\Delta V$ the hatched volume on Fig. \ref{schema_modele}(b).
\begin{align*}
\begin{autobreak}
\Delta V 
= \int_{-\frac{\pi}{2}}^{\frac{\pi}{2}} \diff\theta \int_{0}^{R_{\textsc{\tiny 1/2}}\left( 1 + \left( \frac{H}{2} - \delta(\theta) \right) z'' \right)} \frac{1}{2} \left( r \cos(\theta) \right)^2 z'' r \diff r  
= \frac{\pi}{16} R_{\textsc{\tiny 1/2}}^4 \left( 1 + \left( \frac{H}{2} - \delta(\theta) \right) z'' \right)^4 z'' 
\end{autobreak}
\end{align*}
Or, considering $ R_{\textsc{\tiny 1/2}}  \ll \frac{\lambda^2}{4A} $
\begin{equation*} 
\Delta V = \frac{\pi}{16} R_{\textsc{\tiny 1/2}}^4 z'' 
\end{equation*}
The final expression of cell volume is:
\begin{equation*} 
V=HS_{\textsc{\tiny 1/2}} - \frac{\pi}{16} R_{\textsc{\tiny 1/2}}^4 z''
\end{equation*}

The expressions of the geometric quantities of interest are used in the next section to infer how the cell energy depends on curvature.

\subsection{Energy of the cell}

In this section, the cell energy on a sinusoidal substrate is calculated.
Different terms for the energy are detailed in successive subsections.

The cell shape, i.e. the height and the surface at half height, at equilibrium, are obtained by minimizing the cell energy. 
All energy minimizations are done at constant volume which is equivalent to adding a volume compressibility term $E_v=B\left( V-V_0 \right)^2$ with a very large $B$ modulus.

\subsubsection{Surface energy}

The surface energy of the cell is written 
\begin{equation*}
E_{c} = \gamma_{a}S_{a} + \gamma_{b} S_{b} + \frac{\gamma_{cc}}{2} S_{cc}
\label{energie}
\end{equation*}

The apical, basal and inter-cellular surface energies $\gamma_{a}$, $\gamma_{b}$ and $\gamma_{cc}$ are considered uniform.

On a flat substrate,
\begin{equation*}
E_{c} = \gamma_{a}S_{\textsc{\tiny 1/2}} + \gamma_{b} S_{\textsc{\tiny 1/2}} + \frac{\gamma_{cc}}{2} \alpha\sqrt{S_{\textsc{\tiny 1/2}}}H
\end{equation*}

Since $V = HS_{\textsc{\tiny 1/2}}$, $E_c$ can be written as a function of $S_{\textsc{\tiny 1/2}}$ only :
\begin{equation*}
E_{c} = \gamma_{a}S_{\textsc{\tiny 1/2}} + \gamma_{b} S_{\textsc{\tiny 1/2}} + \frac{\gamma_{cc}}{2} \alpha\frac{V}{\sqrt{S_{\textsc{\tiny 1/2}}}}
\end{equation*}

The minimum surface energy is then obtained for 
\begin{equation*}
S_{\textsc{\tiny 1/2}} = \left( V\frac{\alpha}{4}\frac{\gamma_{cc}}{\gamma_{a} + \gamma_{b}} \right)^{2/3} = \left( V\frac{\alpha}{4\gamma} \right)^{2/3}
\end{equation*}

defining a dimensionless surface tension parameter $\gamma$ :
\begin{equation*}
\gamma=\frac{\gamma_{a} + \gamma_{b}}{\gamma_{cc}}
\end{equation*}

Which can be written equivalently for the height 
\begin{equation*}
H = (V)^{1/3} \left( \frac{4}{\alpha}\gamma \right)^{2/3}
\end{equation*}

If the model with only surface energies holds, the value of the parameter $\gamma$ can be inferred from the measured values of $V$ and $H$ on flat substrate.
\begin{equation*}
\gamma=\frac{\alpha}{4} \sqrt{\frac{H^3}{V}}
\end{equation*}

In the case of a substrate with a sinusoidal profile, the surface energy of a cell on a crest or in a valley is written up to first order in $z''$: 
\begin{align*}
\begin{autobreak}
E_{c} = 
\gamma_{a}S_{\textsc{\tiny 1/2}}\left( 1 - \frac{1}{2}Hz'' \right) 
+ \gamma_{b} S_{\textsc{\tiny 1/2}}\left( 1 + \frac{1}{2}Hz'' \right) 
+ \frac{\gamma_{cc}}{2} \alpha\sqrt{S_{\textsc{\tiny 1/2}}}\left( H-\frac{1}{4}R_{\textsc{\tiny 1/2}}^{2}z'' \right)
\end{autobreak}
\end{align*}

\begin{align*}
\begin{autobreak}
E_{c} =
\gamma_{a}S_{\textsc{\tiny 1/2}}\left( 1 - \frac{1}{2}\frac{V}{S_{\textsc{\tiny 1/2}}}z'' \right) 
+ \gamma_{b} S_{\textsc{\tiny 1/2}}\left( 1 + \frac{1}{2}\frac{V}{S_{\textsc{\tiny 1/2}}}z'' \right) 
+ \frac{\gamma_{cc}}{2} \alpha\sqrt{S_{\textsc{\tiny 1/2}}}\left( \frac{V}{S_{\textsc{\tiny 1/2}}}-\left( \frac{\alpha^2}{16\pi^2} - \frac{\alpha^4}{256\pi^3} \right)S_{\textsc{\tiny 1/2}}z'' \right)
\end{autobreak}
\end{align*}

The relationship $\frac{dEc}{dS_{\textsc{\tiny 1/2}}}=0$ gives a polynomial equation for $S_{\textsc{\tiny 1/2}}$ : 
\begin{align*}
\begin{autobreak}
0 = 
\left( \frac{3\alpha^2}{32\pi}z'' \right)^2 S_{\textsc{\tiny 1/2}}^4 
- \left( \frac{2}{\alpha}\frac{\gamma_{a} 
+ \gamma_{b}}{\frac{\gamma_{cc}}{2}} \right)^2S_{\textsc{\tiny 1/2}}^3 
+ V\frac{3\alpha^2}{16\pi}z''S_{\textsc{\tiny 1/2}}^2 + V^2 
\end{autobreak}
\end{align*}

The analytical expressions of the solutions of this equation are complicated and difficult to handle

The minimization of $E_c(S_{\textsc{\tiny 1/2}})$ can also be performed numerically, after removing the terms that do not depend on $ S_{\textsc{\tiny 1/2}} $ and dividing by $ \alpha \frac{\gamma_{cc}}{2} $ to simplify the expression : 

\begin{equation*}
\frac{E_{c}}{\alpha \frac{\gamma_{cc}}{2}}= - \frac{\alpha^2}{16\pi^2} \left(1 - \frac{\alpha^2}{16\pi} \right) z''S_{\textsc{\tiny 1/2}}^{3/2} + \frac{2}{\alpha}\gamma S_{\textsc{\tiny 1/2}} + \frac{V}{\sqrt{S_{\textsc{\tiny 1/2}}}}
\end{equation*}

The curvature $z''$ of the substrate is equal to zero at the maximum slope of the sinusoid, $z''=0$ also corresponds to the flat case. 
Its maximum (in absolute value) is reached on the crests and in the valleys of the sinusoid. 
The numerical minimization of $E_c$ for these two extreme values of $z''$ gives the values of the cell height on the crests and in the valleys, quantities which were measured in the experiments.

\subsubsection{Line tension}

In this section, the contribution of an \textit{actin belt} near the apical side is discussed by adding a line tension \cite{Hannezo2014} to the the cell energy $E_c$:

\begin{equation*}
E_{c} = \gamma_{a}S_{a} + \gamma_{b} S_{b} + \frac{\gamma_{cc}}{2} S_{cc} +\Lambda_{a}P_a
\end{equation*}

$P_a$ is the apical face perimeter and $\Lambda_a$ its energy cost per unit length.

In the case of a flat substrate, $ P_a = 2\pi R_{\textsc{\tiny 1/2}} $ and 
\begin{equation*}
E_{c} = \gamma_{a}S_{\textsc{\tiny 1/2}} + \gamma_{b} S_{\textsc{\tiny 1/2}} + \frac{\gamma_{cc}}{2} \alpha H \sqrt{S_{\textsc{\tiny 1/2}}} + \Lambda_{a} \alpha \sqrt{S_{\textsc{\tiny 1/2}}} 
\end{equation*}

The minimization of $E_c$ leads to a relationship between the dimensionless surface energy $\gamma$ and the reduced line tension $\Lambda = \frac{\Lambda_a}{\gamma_{cc}}$, which is a length:
\begin{equation} 
\gamma = \frac{\alpha}{4}\frac{1}{\sqrt{S_{\textsc{\tiny 1/2}}}} \left( H - 2\Lambda \right)
\label{annexe_equation_lambda_gamma} 
\end{equation}

The experimental values of $\alpha$, $S_{\textsc{\tiny 1/2}}$ and $H$ give a quantitative relationship between $\gamma$ and $\Lambda$.
In the case of curved substrates, we approximate as previously the perimeter of the apical face by $P_a \approx \pi \left( R_{\textsc{\tiny 1/2}} + R_a \right)$ hence: 
\begin{equation*} 
P \approx 2\pi R_{\textsc{\tiny 1/2}}\left( 1 - \frac{1}{4}Hz'' \right) 
\end{equation*}
The energy of the cell is then written:

\begin{align*}
\begin{autobreak}
E_{c} = 
\gamma_{a}S_{\textsc{\tiny 1/2}}\left( 1 - \frac{1}{2}Hz'' \right) 
+ \gamma_{b} S_{\textsc{\tiny 1/2}}\left( 1 + \frac{1}{2}Hz'' \right) 
+ \frac{\gamma_{cc}}{2} \alpha\sqrt{S_{\textsc{\tiny 1/2}}}\left( H-\frac{1}{4}R_{\textsc{\tiny 1/2}}^{2}z'' \right)
+ 2\pi \Lambda_{a} R_{\textsc{\tiny 1/2}}\left( 1 - \frac{1}{4}Hz'' \right) 
\end{autobreak}
\end{align*}

The cell energy is written as a function of $S_{\textsc{\tiny 1/2}}$ and after removing the terms that do not depend on $S_{\textsc{\tiny 1/2}}$: 

\begin{align*}
\begin{autobreak}
\frac{E_{c}}{\alpha \gamma_{cc}} =
- \frac{\alpha^2}{32\pi^2} \left(1 - \frac{\alpha^2}{16\pi} \right) z''S_{\textsc{\tiny 1/2}}^{3/2} 
+ \frac{1}{\alpha} \gamma S_{\textsc{\tiny 1/2}} 
+ \frac{V}{2\sqrt{S_{\textsc{\tiny 1/2}}}} \left( 1 -  \frac{1}{2} \frac{\Lambda_{a}}{\gamma_{cc}}z'' \right) 
+ \Lambda \sqrt{S_{\textsc{\tiny 1/2}}}
\end{autobreak}
\end{align*}

The numerical minimization of this surface energy is done this time with one adjustable parameter, replacing $\gamma$ by its value as a function of $\Lambda$ obtained from the measurements on flat substrates (Eq \ref{annexe_equation_lambda_gamma}).

\section{Detachment of cells from the substrate at \SI{50}{\micro\Molar} blebbistatin}
\label{detachement}

\begin{figure}[h]
	\begin{center}        
		\includegraphics[width=\columnwidth]{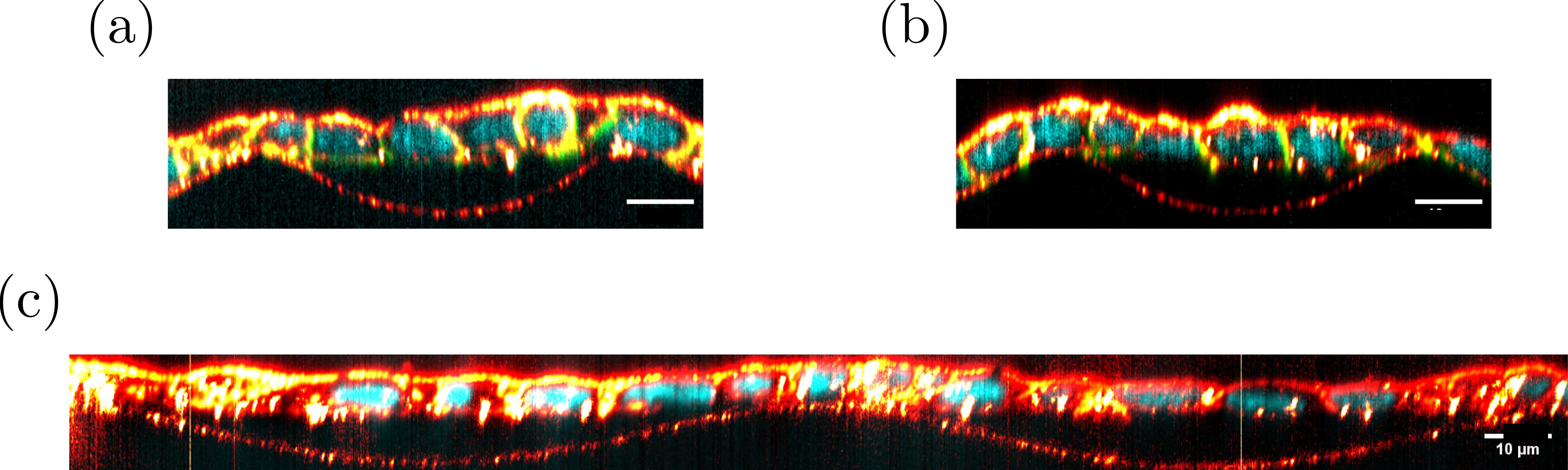}
	\end{center}        
	\caption{MDCK cells exposed to blebbistatin at $\SI{50}{\micro M}$ forming bridges between crests. (a) EcadGFP MDCK cells. (b) WT MDCK cells. Nuclei are in cyan, both F-actin and fibronectin are in red hot, Ecadherin is in green in Fig. a. Scale bars = \SI{10}{\micro\meter}}
        \label{dome_sinus}
\end{figure}

As mentioned in section \ref{myosin_activity}, very few measurements could be made on samples treated with \SI{50}{\micro\Molar} blebbistatin: only 2 for EcadGFP MDCK cells and 1 for WT MDCK cells.
This is due to the fact that in most samples with \SI{50}{\micro\Molar} blebbistatin the epithelium partially detached from curved substrates as illustrated in Figure \ref{dome_sinus}.
This observation is reminiscent of epithelial domes which have been described in several studies \cite{Valentich1979} \cite{Rabito1980} \cite{Lever1979} \cite{latorre_active_2018} and which result from active pumping of fluid through the epithelium.

Detachments mainly occured in the valleys of the sinusoidal substrates making bridges between two successive crests, see Fig. \ref{dome_sinus}. 
Since such bridges are flat, there is no pressure difference between the medium under the basal surface and above the apical surface.
As a consequence the formation of a bridge between two crests can be favorable from an energetic point of view, without, contrary to the case of domes, the need for fluid pumping, provided that:
\[l(\gamma_b + \gamma_a) > l' (\gamma_a + \gamma_b ')\]
with $l$ the length of the portion on substrate from which the epithelium detached and $l'$ the length of the epithelial bridge or the length between the two successive crests.
Here $\gamma_b '$ is the surface tension of the basal face of the cells detached from the substrate.
By assuming $\gamma_b \approx \gamma_b ' - \alpha_b$, with $\alpha_b$ a cell-substrate adhesion energy, defined as positive, the condition for spontaneous epithelium detachment rewrites:
\[\alpha_b < \left(\gamma_a + \gamma_b '\right)\left(1-\frac{l'}{l}\right)\]

At 0 and \SI{5}{\micro\Molar} blebbistatin, the experimental data led to $\gamma_a + \gamma_b < 0 $ , \textit{i.e.} $\alpha_b < (\gamma_a+\gamma'_b)$, which is consistent with the fact that no bridges were observed. 
With \SI{50}{\micro\Molar} blebbistatin on the contrary, the basal adhesion energy probably decreased down to less than $(\gamma_a+\gamma'_b)(1-l'/l)$. 
For instance in Fig. \ref{dome_sinus}(a), $l/l' \approx 1.12$ and $\alpha_b < (\gamma_a+\gamma'_b)/10$.

\begin{acknowledgments}
We would like to thank Andrew Callan-Jones for fruitful discussions, Tom Wyatt for reviewing the manuscript, Alain Richert for his assistance in the laboratory and Christophe Poulard for providing the experimental setup and the protocol to fabricate the sinusoidal profile substrates.
We acknowledge the ImagoSeine core facility of the Institut Jacques Monod (member of the France BioImaging, ANR-10-INBS-04, and the LabEx “Who Am I?” \#ANR-11-LABX-007 (Transition post-doctoral program).
The authors declare no conflict of interest.
\end{acknowledgments}

\bibliographystyle{ieeetr}
\bibliography{Harmand_Henon_3D_shape_of_epithelial_cells_on_curved_substrates}

\begin{thebibliography}{10}

\bibitem{wirtz_physics_2011}
D.~Wirtz, K.~Konstantopoulos, and P.~C. Searson, ``The physics of cancer: The
  role of physical interactions and mechanical forces in metastasis,'' {\em
  Nature Reviews Cancer}, vol.~11, pp.~512--522, July 2011.

\bibitem{hanahan_hallmarks_2011}
D.~Hanahan and R.~A. Weinberg, ``Hallmarks of {{Cancer}}: {{The Next
  Generation}},'' {\em Cell}, vol.~144, pp.~646--674, Mar. 2011.

\bibitem{dasgupta_physics_2018}
S.~Dasgupta, K.~Gupta, Y.~Zhang, V.~Viasnoff, and J.~Prost, ``Physics of lumen
  growth,'' {\em Proceedings of the National Academy of Sciences}, vol.~115,
  pp.~E4751--E4757, May 2018.

\bibitem{mckinley_cellular_2018}
K.~L. McKinley, N.~Stuurman, L.~A. Royer, C.~Schartner, D.~{Castillo-Azofeifa},
  M.~Delling, O.~D. Klein, and R.~D. Vale, ``Cellular aspect ratio and cell
  division mechanics underlie the patterning of cell progeny in diverse
  mammalian epithelia,'' {\em eLife}, vol.~7, p.~e36739, June 2018.

\bibitem{balaji_regulation_2018}
R.~Balaji, V.~Weichselberger, and A.-K. Classen, ``Regulation of tensile stress
  in response to external forces coordinates epithelial cell shape transitions
  with organ growth and elongation,'' {\em bioRxiv}, p.~399303, Aug. 2018.

\bibitem{Monier2015}
B.~Monier, M.~Gettings, G.~Gay, T.~Mangeat, S.~Schott, A.~Guarner, and
  M.~Suzanne, ``Apico-basal forces exerted by apoptotic cells drive epithelium
  folding.,'' {\em Nature}, vol.~518, no.~7538, 2015.

\bibitem{fouchard_curling_2020}
J.~Fouchard, T.~P.~J. Wyatt, A.~Proag, A.~Lisica, N.~Khalilgharibi, P.~Recho,
  M.~Suzanne, A.~Kabla, and G.~Charras, ``Curling of epithelial monolayers
  reveals coupling between active bending and tissue tension,'' {\em
  Proceedings of the National Academy of Sciences}, Apr. 2020.

\bibitem{xi_emergent_2017}
W.~Xi, S.~Sonam, T.~B. Saw, B.~Ladoux, and C.~T. Lim, ``Emergent patterns of
  collective cell migration under tubular confinement,'' {\em Nature
  Communications}, vol.~8, pp.~1--15, Nov. 2017.

\bibitem{yevick_architecture_2015}
H.~G. Yevick, G.~Duclos, I.~Bonnet, and P.~Silberzan, ``Architecture and
  migration of an epithelium on a cylindrical wire,'' {\em Proceedings of the
  National Academy of Sciences}, vol.~112, pp.~5944--5949, May 2015.

\bibitem{Thompson1945}
D.~W. Thompson, {\em On Growth and Form}.
\newblock 1945.

\bibitem{steinberg_reconstruction_1963}
M.~S. Steinberg, ``Reconstruction of {{Tissues}} by {{Dissociated Cells}},''
  {\em Science}, vol.~141, pp.~401--408, Aug. 1963.

\bibitem{farhadifar_influence_2007}
R.~Farhadifar, J.-C. R{\"o}per, B.~Aigouy, S.~Eaton, and F.~J{\"u}licher, ``The
  influence of cell mechanics, cell-cell interactions, and proliferation on
  epithelial packing,'' {\em Current biology: CB}, vol.~17, pp.~2095--2104,
  Dec. 2007.

\bibitem{landsberg_increased_2009}
K.~P. Landsberg, R.~Farhadifar, J.~Ranft, D.~Umetsu, T.~J. Widmann, T.~Bittig,
  A.~Said, F.~J{\"u}licher, and C.~Dahmann, ``Increased {{Cell Bond Tension
  Governs Cell Sorting}} at the {{Drosophila Anteroposterior Compartment
  Boundary}},'' {\em Current Biology}, vol.~19, pp.~1950--1955, Dec. 2009.

\bibitem{aegerter-wilmsen_integrating_2012}
T.~{Aegerter-Wilmsen}, M.~B. Heimlicher, A.~C. Smith, P.~B. de~Reuille, R.~S.
  Smith, C.~M. Aegerter, and K.~Basler, ``Integrating force-sensing and
  signaling pathways in a model for the regulation of wing imaginal disc
  size,'' {\em Development}, vol.~139, pp.~3221--3231, Sept. 2012.

\bibitem{Bi2015}
D.~Bi, J.~H. Lopez, J.~M. Schwarz, and M.~L. Manning, ``A density-independent
  rigidity transition in biological tissues,'' {\em Nature Physics}, vol.~11,
  no.~12, 2015.

\bibitem{Wen2017}
F.~L. Wen, Y.~C. Wang, and T.~Shibata, ``Epithelial {{Folding Driven}} by
  {{Apical}} or {{Basal}}-{{Lateral Modulation}}: {{Geometric Features}},
  {{Mechanical Inference}}, and {{Boundary Effects}},'' {\em Biophysical
  Journal}, vol.~112, no.~12, pp.~2683--2695, 2017.

\bibitem{polyakov_passive_2014}
O.~Polyakov, B.~He, M.~Swan, J.~W. Shaevitz, M.~Kaschube, and E.~Wieschaus,
  ``Passive {{Mechanical Forces Control Cell}}-{{Shape Change}} during
  {{Drosophila Ventral Furrow Formation}},'' {\em Biophysical Journal},
  vol.~107, pp.~998--1010, Aug. 2014.

\bibitem{Storgel2016}
N.~{\v S}torgel, M.~Krajnc, P.~Mrak, J.~{\v S}trus, and P.~Ziherl,
  ``Quantitative {{Morphology}} of {{Epithelial Folds}},'' {\em Biophysical
  Journal}, vol.~110, no.~1, pp.~269--277, 2016.

\bibitem{Hannezo2014}
E.~Hannezo, J.~Prost, and J.-F. Joanny, ``Theory of epithelial sheet morphology
  in three dimensions.,'' {\em Proceedings of the National Academy of
  Sciences}, vol.~111, no.~1, pp.~27--32, 2014.

\bibitem{honda_three-dimensional_2004}
H.~Honda, M.~Tanemura, and T.~Nagai, ``A three-dimensional vertex dynamics cell
  model of space-filling polyhedra simulating cell behavior in a cell
  aggregate,'' {\em Journal of Theoretical Biology}, vol.~226, pp.~439--453,
  Feb. 2004.

\bibitem{Misra2016}
M.~Misra, B.~Audoly, I.~G. Kevrekidis, and S.~Y. Shvartsman, ``Shape
  {{Transformations}} of {{Epithelial Shells}},'' {\em Biophysical Journal},
  vol.~110, no.~7, pp.~1670--1678, 2016.

\bibitem{Dukes2011}
J.~D. Dukes, P.~Whitley, and A.~D. Chalmers, ``The {{MDCK}} variety pack:
  Choosing the right strain,'' {\em BMC Cell Biology}, vol.~12, no.~1, p.~43,
  2011.

\bibitem{graner_equilibrium_2000}
F.~Graner, Y.~Jiang, E.~Janiaud, and C.~Flament, ``Equilibrium states and
  ground state of two-dimensional fluid foams,'' {\em Physical Review E},
  vol.~63, p.~011402, Dec. 2000.

\bibitem{bi_motility-driven_2016}
D.~Bi, X.~Yang, M.~C. Marchetti, and M.~L. Manning, ``Motility-{{Driven Glass}}
  and {{Jamming Transitions}} in {{Biological Tissues}},'' {\em Physical Review
  X}, vol.~6, Apr. 2016.

\bibitem{bambardekar_direct_2015}
K.~Bambardekar, R.~Cl{\'e}ment, O.~Blanc, C.~Chard{\`e}s, and P.-F. Lenne,
  ``Direct laser manipulation reveals the mechanics of cell contacts in
  vivo.,'' {\em Proceedings of the National Academy of Sciences of the United
  States of America}, vol.~112, pp.~1416--21, Feb. 2015.

\bibitem{salbreux_actin_2012}
G.~Salbreux, G.~Charras, and E.~Paluch, ``Actin cortex mechanics and cellular
  morphogenesis,'' {\em Trends in Cell Biology}, vol.~22, pp.~536--545, Oct.
  2012.

\bibitem{solon_pulsed_2009}
J.~Solon, A.~{Kaya-{\c C}opur}, J.~Colombelli, and D.~Brunner, ``Pulsed
  {{Forces Timed}} by a {{Ratchet}}-like {{Mechanism Drive Directed Tissue
  Movement}} during {{Dorsal Closure}},'' {\em Cell}, vol.~137, pp.~1331--1342,
  June 2009.

\bibitem{Maitre2012}
J.~L. Maitre, H.~Berthoumieux, S.~F.~G. Krens, G.~Salbreux, F.~Julicher,
  E.~Paluch, and C.~P. Heisenberg, ``Adhesion {{Functions}} in {{Cell Sorting}}
  by {{Mechanically Coupling}} the {{Cortices}} of {{Adhering Cells}},'' {\em
  Science}, vol.~338, pp.~253--256, Oct. 2012.

\bibitem{adams_mechanisms_1998}
C.~L. Adams, Y.-T. Chen, S.~J. Smith, and W.~James~Nelson, ``Mechanisms of
  {{Epithelial Cell}}\textendash{{Cell Adhesion}} and {{Cell Compaction
  Revealed}} by {{High}}-resolution {{Tracking}} of
  {{E}}-{{Cadherin}}\textendash\,{{Green Fluorescent Protein}},'' {\em The
  Journal of Cell Biology}, vol.~142, pp.~1105--1119, Aug. 1998.

\bibitem{Foty2005a}
R.~A. Foty and M.~S. Steinberg, ``The differential adhesion hypothesis: {{A}}
  direct evaluation,'' {\em Developmental Biology}, vol.~278, no.~1,
  pp.~255--263, 2005.

\bibitem{maitre_adhesion_2012}
J.-L. Ma{\^i}tre, H.~Berthoumieux, S.~F.~G. Krens, G.~Salbreux,
  F.~J{\"u}licher, E.~Paluch, and C.-P. Heisenberg, ``Adhesion {{Functions}} in
  {{Cell Sorting}} by {{Mechanically Coupling}} the {{Cortices}} of {{Adhering
  Cells}},'' {\em Science}, vol.~338, pp.~253--256, Oct. 2012.

\bibitem{Murrell2015}
M.~Murrell, P.~W. Oakes, M.~Lenz, and M.~L. Gardel, ``Forcing cells into shape:
  The mechanics of actomyosin contractility.,'' {\em Nature reviews. Molecular
  cell biology}, vol.~16, no.~8, pp.~486--498, 2015.

\bibitem{harris_adherens_2010}
T.~J.~C. Harris and U.~Tepass, ``Adherens junctions: From molecules to
  morphogenesis,'' {\em Nature Reviews Molecular Cell Biology}, vol.~11,
  pp.~502--514, July 2010.

\bibitem{mitrossilis_single-cell_2009}
D.~Mitrossilis, J.~Fouchard, A.~Guiroy, N.~Desprat, N.~Rodriguez, B.~Fabry, and
  A.~Asnacios, ``Single-cell response to stiffness exhibits muscle-like
  behavior,'' {\em Proceedings of the National Academy of Sciences}, vol.~106,
  pp.~18243--18248, Oct. 2009.

\bibitem{wyatt_question_2016}
T.~Wyatt, B.~Baum, and G.~Charras, ``A question of time: Tissue adaptation to
  mechanical forces,'' {\em Current Opinion in Cell Biology}, vol.~38,
  pp.~68--73, Feb. 2016.

\bibitem{liu_mechanical_2010}
Z.~Liu, J.~L. Tan, D.~M. Cohen, M.~T. Yang, N.~J. Sniadecki, S.~A. Ruiz, C.~M.
  Nelson, and C.~S. Chen, ``Mechanical tugging force regulates the size of
  cell\textendash{}cell junctions,'' {\em Proceedings of the National Academy
  of Sciences}, vol.~107, pp.~9944--9949, June 2010.

\bibitem{furukawa_epithelial_2017}
K.~T. Furukawa, K.~Yamashita, N.~Sakurai, and S.~Ohno, ``The {{Epithelial
  Circumferential Actin Belt Regulates YAP}}/{{TAZ}} through
  {{Nucleocytoplasmic Shuttling}} of {{Merlin}},'' {\em Cell Reports}, vol.~20,
  pp.~1435--1447, Aug. 2017.

\bibitem{uyeda_stretching_2011}
T.~Q.~P. Uyeda, Y.~Iwadate, N.~Umeki, A.~Nagasaki, and S.~Yumura, ``Stretching
  {{Actin Filaments}} within {{Cells Enhances}} their {{Affinity}} for the
  {{Myosin II Motor Domain}},'' {\em PLoS ONE}, vol.~6, Oct. 2011.

\bibitem{herzlinger_mdck_1982}
D.~A. Herzlinger, T.~G. Easton, and G.~K. Ojakian, ``The {{MDCK}} epithelial
  cell line expresses a cell surface antigen of the kidney distal tubule,''
  {\em The Journal of Cell Biology}, vol.~93, pp.~269--277, May 1982.

\bibitem{sage_mij_2012}
D.~Sage, D.~Prodanov, J.-Y. Tinevez, and J.~Schindelin, ``{{MIJ}}: {{Making
  Interoperability Between ImageJ}} and {{Matlab Possible}},'' p.~1, 2012.

\bibitem{bowden_spontaneous_1998}
N.~Bowden, S.~Brittain, A.~G. Evans, J.~W. Hutchinson, and G.~M. Whitesides,
  ``Spontaneous formation of ordered structures in thin films of metals
  supported on an elastomeric polymer,'' {\em Nature}, vol.~393, pp.~146--149,
  May 1998.

\bibitem{brau_multiple-length-scale_2011}
F.~Brau, H.~Vandeparre, A.~Sabbah, C.~Poulard, A.~Boudaoud, and P.~Damman,
  ``Multiple-length-scale elastic instability mimics parametric resonance of
  nonlinear oscillators,'' {\em Nature Physics}, vol.~7, pp.~56--60, Jan. 2011.

\bibitem{Valentich1979}
J.~Valentich, ``Hemicyst formation stimulated by cyclic {{AMP}} in dog kidney
  cell line {{MDCK}},'' {\em J. Cell. Physiol.}, vol.~100, pp.~291--304, Aug.
  1979.

\bibitem{Rabito1980}
C.~A. Rabito, R.~Tchao, J.~Valentich, and J.~Leighton, ``Effect of
  cell-substratum interaction on hemicyst formation by {{MDCK}} cells,'' {\em
  In Vitro}, vol.~16, no.~6, pp.~461--468, 1980.

\bibitem{Lever1979}
J.~E. Lever, ``Regulation of dome formation in differentiated epithelial cell
  cultures.,'' {\em Journal of supramolecular structure}, vol.~12,
  pp.~259--272, 1979.

\bibitem{latorre_active_2018}
E.~Latorre, S.~Kale, L.~Casares, M.~{G{\'o}mez-Gonz{\'a}lez}, M.~Uroz,
  L.~Valon, R.~V. Nair, E.~Garreta, N.~Montserrat, A.~del Campo, B.~Ladoux,
  M.~Arroyo, and X.~Trepat, ``Active superelasticity in three-dimensional
  epithelia of controlled shape,'' {\em Nature}, vol.~563, pp.~203--208, Nov.
  2018.

\end{thebibliography}

\end{document}